\documentclass[prb,aps,twocolumn,superscriptaddress,showpacs]{revtex4}
\usepackage{epsfig}
\usepackage{amsmath}
\usepackage{amsfonts}
\usepackage{amssymb,mathrsfs}
\usepackage{esint}

\newcommand{\bs}[1]{\boldsymbol{#1}}

\begin{document}

\title{Hydrodynamics in graphene: Linear-response transport}

\author{B.N. Narozhny} 
\affiliation{Institut f\"ur Theorie der Kondensierten Materie,
  Karlsruher Institut f\"ur Technologie, 76128 Karlsruhe, Germany}
\affiliation{National Research Nuclear University MEPhI (Moscow Engineering Physics Institute), 
  Kashirskoe shosse 31, 115409 Moscow, Russia }

\author{I.V. Gornyi} 
\affiliation{Institut f\"ur Nanotechnologie, Karlsruhe Institute of
  Technology, 76021 Karlsruhe, Germany}
\affiliation{A.F. Ioffe Physico-Technical Institute, 194021
  St. Petersburg, Russia}

\author{M. Titov} 
\affiliation{Radboud University Nijmegen,
  Institute for Molecules and Materials, NL-6525 AJ Nijmegen, The
  Netherlands}

\author{M. Sch\"utt}
\affiliation{School of Physics and Astronomy, University of Minnesota, Minneapolis, MN 55455, USA}

\author{A.D. Mirlin}
\affiliation{Institut f\"ur Nanotechnologie, Karlsruhe Institute of Technology,
 76021 Karlsruhe, Germany}
\affiliation{Institut f\"ur Theorie der Kondensierten Materie,
  Karlsruher Institut f\"ur Technologie, 76128 Karlsruhe, Germany}
\affiliation{Petersburg Nuclear Physics Institute,
 188350 St. Petersburg, Russia}

\date{\today}

\begin{abstract}
  We develop a hydrodynamic description of transport properties in
  graphene-based systems which we derive from the quantum kinetic
  equation. In the interaction-dominated regime, the collinear
  scattering singularity in the collision integral leads to fast
  unidirectional thermalization and allows us to describe the system
  in terms of three macroscopic currents carrying electric charge,
  energy, and quasiparticle imbalance. Within this ``three-mode''
  approximation we evaluate transport coefficients in monolayer
  graphene as well as in double-layer graphene-based structures. The
  resulting classical magnetoresistance is strongly sensitive to the
  interplay between the sample geometry and leading relaxation
  processes. In small, mesoscopic samples the macroscopic currents are
  inhomogeneous which leads to linear magnetoresistance in classically
  strong fields. Applying our theory to double-layer graphene-based
  systems, we provide microscopic foundation for phenomenological
  description of giant magnetodrag at charge neutrality and find
  magnetodrag and Hall drag in doped graphene.
\end{abstract}


\maketitle

Traditional hydrodynamics \cite{dau} describes systems at large length
scales (compared to the mean free path). The hydrodynamic equations
are typically formulated in terms of currents and densities of
conserved quantities and can be derived from the kinetic equation
using either the Chapman-Enskog \cite{cen} or Grad \cite{gra}
procedures. Within the leading approximation, gradients of the
macroscopic physical quantities are assumed to be small, such that the
system can be characterized by the {\it local equilibrium}
distribution function. Dissipative properties, such as electrical or
thermal conductivity or viscosity are then determined by small
corrections to the local-equilibrium distribution function. Within
linear response, such corrections are proportional to a weak external
bias.

Recently the kinetic equation approach was applied to electronic
excitations in graphene \cite{kas,mu1,mu2,kin,fos,ryz,mem}. In contrast to
conventional metals and semiconductors, graphene is characterized by
the linear excitation spectrum which makes the system explicitly {\it
  non-Galilean-invariant}. Consequently, the transport scattering time
in graphene is strongly affected by electron-electron interaction
\cite{pol} which has to be taken into account on equal footing with
disorder potential. At the same time, due to the classical nature of
the Coulomb interaction between charge carriers in graphene, the
system is also {\it non-Lorentz-invariant}. As a result, the standard
derivation \cite{dau} of the hydrodynamic equations from the kinetic
equation has to be revisited \cite{kas,mu1,mu2,kin,fos,ryz,mem}.

The linearity of the quasiparticle spectrum in graphene leads to an
important corollary: the energy and momentum conservation laws for
Dirac quasiparticles coincide in the special case of collinear
scattering. This kinematic peculiarity results in a singular
contribution to the collision integral \cite{kas,mu1,kin,mem} allowing
for a non-perturbative solution to the kinetic equation. The distinct
feature of this solution is fast unidirectional thermalization
\cite{mem} that facilitates integration of the kinetic equation. The
unique feature of the resulting hydrodynamic description of electronic
transport in graphene is inequivalence of the electric current and
total momentum of the system \cite{mu1,mu2,mem}. As the latter is
equivalent to the energy current, transport properties of graphene are
governed by a non-trivial interplay of electric current and energy
relaxation.

Two-fluid hydrodynamics in graphene was suggested in
Refs.~\onlinecite{mu1,fos} and then extended to double-layer
graphene-based structures in Ref.~\onlinecite{mem}, which allowed for
a description of the Coulomb drag effect \cite{gor,tu1,tu2,meg} in
graphene. An extension of this approach to mesoscopic (finite-size)
samples was suggested in Ref.~\onlinecite{meg}. Qualitatively, this
theory can be interpreted in terms of a semiclassical {\it two-band}
model that yields non-trivial magnetic field dependence of the
transport coefficients and accounts for the effect of {\it giant
  magnetodrag} at the neutrality point \cite{meg}. The classical
mechanism of this effect is similar to the standard mechanism of
magnetoresistance in multi-band systems \cite{bsm}.

In this paper we rigorously derive the hydrodynamic description of
electronic transport in graphene within linear response. While we use
the same collinear-scattering singularity as found in
Refs.~\onlinecite{kas,mu1,kin,mem} in order to integrate the quantum
kinetic equation, we argue that the physics of the system should be
described in terms of {\it three} macroscopic currents: the electric
current $\bs{j}$, energy current $\bs{Q}$, and quasiparticle imbalance
\cite{fos} current $\bs{P}$. 

For general doping, the resulting theory is rather cumbersome.
However, at the charge neutrality point and in the degenerate limit
the equations simplify allowing for an analytic solution. In the
former case, we focus on the issue of magnetoresistance, a subject of
considerable experimental interest
\cite{fu1,fu2,job,mg1,mg2,mg3,mg4}. In particular, we demonstrate the
appearance of the linear magnetoresistance in moderately strong,
classical magnetic fields in monolayer graphene \cite{mrs}. In
double-layer graphene-based systems we describe {\it negative} Coulomb
drag \cite{meg,lev} and justify the phenomenological two-band model of
Ref.~\onlinecite{meg} (precisely at the Dirac point the imbalance
current is proportional to the energy current allowing one to reduce
the number of variables). Both effects occur in narrow, mesoscopic
samples in the presence of energy relaxation and quasiparticle
recombination due to electron-phonon interaction.

In the opposite limit of very high doping (i.e. in the ``Fermi-liquid
regime'') all three macroscopic currents become equivalent and the
theory is reduced to the standard Drude-like description that can be
also derived by perturbative methods \cite{us1}. Here we find the
leading corrections to the standard picture of Coulomb drag
\cite{us1,thd} yielding magnetodrag and Hall drag in doped graphene.

The rest of the paper is organized as follows. We begin
(Section~\ref{sum}) with the summary of our theory and results for
monolayer graphene. In Section~\ref{hdm}, we present a derivation of
the macroscopic description of electronic transport. In
Section~\ref{fse}, we use this theory to evaluate transport
coefficients in graphene such as the magnetoresistance at the point of
charge neutrality for small, mesoscopic samples. In Section~\ref{dlg},
we apply our theory to double-layer graphene-based systems
\cite{gor,tu1,tu2,meg}. Concluding remarks can be found in
Section~\ref{disc}. Technical details are relegated to the Appendices.


\section{Macroscopic description of transport in monolayer graphene}
\label{sum}

In this Section, we describe transport properties of monolayer
graphene. Neglecting all quantum effects \cite{aar,zna,fal}, we base our
considerations on the set of macroscopic transport equations which
essentially generalize the usual Ohm's law to the case of
collision-dominated transport in graphene. These equations can be
derived from the kinetic equation (see Section~\ref{hdm} below) in the
interaction-dominated regime, where the transport scattering time due
to electron-electron interaction $\tau_{ee}$ is much smaller than the
disorder mean-free time $\tau$
\[
\tau_{ee}\ll\tau.
\]
We limit ourselves to the discussion of a steady state. The latter is
typically established by means of disorder scattering. A notable
exception is neutral graphene in the absence of magnetic field, where
the steady state exist due to electron-electron interaction alone.
However, in the presence of the field, even at the Dirac point the
steady state cannot be reached without disorder. Therefore, we have
to keep the weak disorder in the problem. For simplicity, we assume
the mean-free time $\tau$ to be energy independent, although in
physical graphene most of the impurity scattering processes lead to
energy-dependent relaxation rates. A corresponding generalization of
our theory is straightforward \cite{us1} and does not lead to
qualitatively new effects \cite{fn1}. At the same time, quantitative description
of experimental data may greatly benefit from a realistic description
of disorder \cite{meg}.

Non-linear hydrodynamics of graphene will be discussed in a separate
publication \cite{ulf}.

\subsection{Linear response equations in graphene}

One of the main results of this paper is the set of macroscopic
equations describing electronic transport in graphene within linear
response. What makes this unusual is that the electric current
$\bs{j}$ is inequivalent to the energy current $\bs{Q}$ and the
quasiparticle imbalance current $\bs{P}$. The three macroscopic
currents can be found from the following equations
\begin{subequations}
\label{meqs}
\begin{equation}
\label{jeq}
-\bs{\nabla}\Pi +  \bs{E} + {\cal R}_H\bs{\cal K}\times \bs{e}_{\bs{B}}
= {\cal R}_0 \bs{j} + \frac{\pi}{e^2 K}\left[\frac{\bs{\cal A}}{\tau_{vv}} +
\frac{\bs{\cal C}}{\tau_{vs}}\right],
\end{equation}
\begin{eqnarray}
\label{qeq}
-\bs{\nabla}\Theta +{\cal N}_1\bs{E} + {\cal R}_H\left[ \bs{j}\times \bs{e}_{\bs{B}}\right]
= {\cal R}_0 \frac{e}{K}\bs{Q},
\end{eqnarray}
\begin{equation}
\label{peq}
-\bs{\nabla}\Xi +\frac{\mu}{K}\bs{E} 
+ {\cal R}_H\widetilde{\bs{\cal K}}\times \bs{e}_{\bs{B}}
= e {\cal R}_0 \bs{P} 
+ \frac{\pi}{e^2 K}\left[\frac{\bs{\cal A}}{\tau_{vs}} +
\frac{\bs{\cal C}}{\tau_{ss}}\right].
\end{equation}
\end{subequations}
Here $\bs{E}$ is the electric field, $\bs{e}_{\bs{B}}$ is the unit
vector in the direction of the magnetic field
$\bs{B}=B\bs{e}_{\bs{B}}$, $K$ is the mean quasiparticle kinetic
energy in graphene \cite{meg} (with $T$ being the temperature and
$\mu$ the chemical potential):
\begin{equation}
\label{kt}
K = 2T\ln\left(1+e^{\mu/T}\right)-\mu
\rightarrow
\left\{
\begin{matrix}
\mu & , & T\ll\mu \cr
2T\ln 2 & , & T\gg\mu
\end{matrix}
\right.,
\end{equation}
the dimensionless quantity ${{\cal N}_1=2n_0/(\nu_0 K)}$ represents
the equilibrium charge density $n_0$ (here $\nu_0=\partial
n_0/\partial\mu$), and the two coefficients ${\cal R}_0$ and ${\cal
  R}_H$ are
\begin{equation}
\label{r0rh}
{\cal R}_0(\mu, \tau, T) = \frac{\pi}{e^2 K\tau}, \quad
{\cal R}_H(\mu, B, T) = \frac{\pi\omega_B}{e^2K},
\end{equation}
where $e$ the electron charge, the frequency $\omega_B$ is
\begin{equation}
\label{cfr}
\omega_B = \frac{ev_g^2B}{cK},
\end{equation}
and $v_g$ the quasiparticle velocity.

In graphene, the energy current $\bs{Q}$ is equivalent to the total
momentum of electrons, which cannot be relaxed by electron-electron
interaction respecting momentum conservation. Therefore, the transport
scattering rates due to electron-electron interaction appear only in
Eqs.~(\ref{jeq}) and (\ref{peq}). The three scattering times
$\tau_{vv}$, $\tau_{vs}$, and $\tau_{ss}$ describe mutual scattering
of the velocity and imbalance modes respecting Onsager reciprocity.

The above three modes form the {\it three-mode Ansatz} for the 
non-equilibrium correction to the electronic distribution function
[see Eq.~(\ref{df}) below and Appendix~\ref{Ah}]
\begin{equation}
\label{habc}
h = 
\frac{2\bs{v}}{e \nu_0 T v_g^2}
\left\{\bs{\cal A} + \bs{\cal B}\frac{\epsilon}{K} 
+ \bs{\cal C}{\rm sign}(\epsilon)\right\},
\end{equation}
where the vectors $\bs{\cal A}$, $\bs{\cal B}$, and $\bs{\cal C}$ are
linear combinations of the three macroscopic currents that are
introduced for brevity [see Eq.~(\ref{h3m}) for details]. The absence
of the vector $\bs{\cal B}$ in the right-hand side of
Eqs.~(\ref{meqs}) is due to momentum conservation. However, all three
auxiliary vectors enter the Lorentz terms in the following
combinations
\begin{equation}
\label{ka}
\bs{\cal K} = \bs{\cal A} \tanh\frac{\mu}{2T} + \bs{\cal B} + \bs{\cal C},
\end{equation}
\begin{eqnarray}
\label{kta}
\widetilde{\bs{\cal K}} = \bs{\cal A} + \bs{\cal B}\frac{\mu}{K} 
+ \bs{\cal C} \tanh\frac{\mu}{2T}.
\end{eqnarray}

The quantity $\Pi$ represents the inhomogeneous part of the flux
density of the electric current (cf. the usual momentum flux density
or the ``stress-tensor'') and is given by a linear combination of the
inhomogeneous densities corresponding to the three modes in the
system: the charge $\delta n$, energy $\delta u$, and imbalance
$\delta\rho$. Similarly, the quantities $\Theta$ and $\Xi$ describe
the flux densities for the energy and imbalance currents, see
Eqs.~(\ref{grs}) below.

In finite-size samples the equations (\ref{meqs}) have to be
supplemented by the corresponding continuity equations and Maxwell's
equations, since inhomogeneous charge density fluctuations give rise
to electromagnetic fields. Therefore the electric field $\bs{E}$ in
Eqs.~(\ref{meqs}) comprises the externally applied and self-consistent
(Vlasov-like \cite{dau}) fields. The self-consistency amounts to
solving the electrostatic problem described by the Maxwell's equations
\cite{dau}
\begin{equation}
\label{max}
\bs{\nabla}\cdot\bs{E} = 4\pi \delta n \delta(z), \quad \bs{\nabla}\times\bs{E} = 0, \quad
\bs{\nabla}\times\bs{B} = \frac{4\pi}{c} \bs{j}.
\end{equation}
While charge carriers are confined within the graphene
sheet, the electromagnetic fields are not, hence the factor of
$\delta(z)$ in Eq.~(\ref{max}). At the same time, we assume that the
uniform charge density $n_0$ is controlled by an external
gate. Consequently, only the non-uniform part of the charge density
$\delta n$ is taken into account in Eq.~(\ref{max}).

The continuity equations can be obtained by integrating the kinetic
equation in the usual fashion \cite{dau}. In the steady state, charge
conservation requires
\begin{subequations}
\label{ceqsdp}
\begin{equation}
\label{jceqdp}
\bs{\nabla}\cdot\bs{j} = 0.
\end{equation}
Similar equations can be derived for the energy and imbalance
density. Since both of them are conserved by electron-electron
interactions, the collision integral in Eq.~(\ref{beq0}) does not
contribute to the continuity equations. At the same time,
electron-phonon interaction (that we have so far neglected) may lead
to energy and imbalance relaxation
processes\cite{fos,meg,hwa,fra,bis,kub,tse,vil,son}. Taking into
account the electron-phonon collisions, we find the following
continuity equations (see Appendix~\ref{eph} for details):
\begin{equation}
\label{iceqdp}
e\bs{\nabla}\cdot\bs{P} = -\frac{\mathfrak{b}}{\tau_{Ib}} 
+ \frac{\mathfrak{c}}{\tau_{Ic}},
\end{equation}
\begin{equation}
\label{eceqdp}
\frac{e}{K}\bs{\nabla}\cdot\bs{Q} = \frac{\mathfrak{b}}{\tau_{Eb}} 
- \frac{\mathfrak{c}}{\tau_{Ec}}.
\end{equation}
\end{subequations}
Here the auxiliary quantities $\mathfrak{b}$ and $\mathfrak{c}$ are
linear combinations of inhomogeneous parts of the charge, energy, and
imbalance densities with the same coefficients as the vectors
$\bs{\cal B}$ and $\bs{\cal C}$, see Eq.~(\ref{dh3m}). Physically,
imbalance relaxation (described by $\tau_{Ib}$ and $\tau_{Ic}$) is due
to inter-band processes only and thus is expected to be slower than
energy relaxation (described by $\tau_{Eb}$ and $\tau_{Ec}$).

The macroscopic equations (\ref{meqs}) simplify at the neutrality
point and in the degenerate (or Fermi-liquid) limit. We now turn to
the discussion of the solutions to Eqs.~(\ref{meqs}) in these cases,
which clarify the structure of our theory.

\subsection{Transport in the degenerate limit}

At high doping (or at low temperatures), the electronic system in
graphene becomes degenerate. In the limit $\mu\gg T$, all three
macroscopic currents become equivalent
\begin{equation}
\label{flcs}
\bs{j}(\mu\gg T) \approx \frac{e}{\mu} \bs{Q}(\mu\gg T) \approx e\bs{P} (\mu\gg T).
\end{equation}
The additional vectors introduced in Eqs.~(\ref{meqs}) simplify to
\[
\bs{\cal A}(\mu\gg T)\approx\bs{\cal K}(\mu\gg T)
\approx\widetilde{\bs{\cal K}}(\mu\gg T)\approx\bs{j},
\]
\[
\bs{\cal C}(\mu\gg T)=0.
\]
In this regime, the Galilean invariance is effectively restored and
all relaxation rates due to electron-electron interaction vanish.
Consequently, the three equations (\ref{meqs}) become equivalent to
the Ohm's law 
\begin{equation}
\label{drude-ms}
-\frac{\bs{\nabla} \delta n}{e^2\nu_0}  + 
\bs{E} + {R}_H \bs{j}\times\bs{e}_{\bs{B}} = {R}_0\bs{j},
\end{equation}
where
\begin{subequations}
\label{flr}
\begin{equation}
\label{r0fl}
{R}_0 = {\cal R}_0(\mu\gg T)=\frac{\pi}{e^2\mu\tau},
\end{equation}
and
\begin{equation}
\label{rhfl}
{R}_H = {\cal R}_H(\mu\gg T)=\frac{\pi v_g^2 B}{ec\mu^2},
\end{equation}
\end{subequations}
are the usual longitudinal and Hall resistances.

Physically, the above simplification is related to the fact, that in
the degenerate regime inter-band processes are exponentially
suppressed. Effectively only one band participates in transport and
therefore the textbook results apply; in particular there is no
magnetoresistance. For leading corrections to this behavior see
Section~\ref{cddl}.

\subsection{Transport at the neutrality point}

At the charge neutrality point $\mu=0$, the auxiliary vectors in
Eqs.~(\ref{meqs}) have the form
\begin{subequations}
\label{aim0}
\begin{eqnarray}
&&
\bs{\cal A}(\mu=0) = \widetilde{\bs{\cal K}}(\mu=0) = \bs{j}, 
\\
&&
\nonumber\\
&&
\bs{\cal C}(\mu=0) = \gamma_0 \frac{e\bs{Q}}{2T\Delta(0)\ln 2}
- e\bs{P}\frac{{\cal N}_2(0)}{\Delta(0)}, 
\nonumber\\
&&
\nonumber\\
&&
\bs{\cal K}(\mu=0)=
\frac{\gamma_0-1}{\Delta(0)}\left[\frac{e\bs{Q}}{2T\ln 2} - e\bs{P}\gamma_2\right].
\nonumber
\end{eqnarray}
Here the numerical coefficients are
\begin{equation}
\label{g0}
\gamma_0 = \pi^2/(12\ln^22) \approx 1.7119, 
\end{equation}
\begin{equation}
\label{n20}
{\cal N}_2(0) = 9\zeta(3)/(8\ln^32)\approx 4.0607,
\end{equation}
\begin{equation}
\label{g2}
\gamma_2 = ({\cal N}_2(0) - \gamma_0)/(\gamma_0-1)
\approx 3.2996,
\end{equation}
and
\begin{equation}
\label{d0}
\Delta(0) = \gamma_0^2 - {\cal N}_2(0) \approx -1.1303.
\end{equation}
\end{subequations}
In addition, one of the relaxation rates vanishes as well
\[
\tau_{vs}^{-1}(\mu=0)=0.
\]
As a result, the equations (\ref{meqs}) simplify. Below we consider
the two limiting cases of wide and narrow samples as determined by the
interplay between the electron-phonon scattering and the magnetic field
\cite{mrs}.

\subsubsection{Transport coefficients in macroscopic samples}

If the sample width is the largest length scale in the problem,
$W\gg\ell_R\omega_B^2\tau\tau_{ee}$ (where $\tau_{ee}$ is the typical
value of the electron-electron transport scattering times and $\ell_R$
is the typical length scale describing quasiparticle recombination due
to electron-phonon scattering, see Section~\ref{slmr}), the boundary
effects may be neglected and the sample behaves as if it were
infinite. Then all physical quantities can be considered uniform. At
charge neutrality, the equations (\ref{meqs}) take the form
\begin{subequations}
\label{sleq3dp}
\begin{equation}
\label{sljeq3dp}
\bs{E} + {\cal R}_H\bs{\cal K}\times \bs{e}_{\bs{B}}
= {\cal R}_0 \bs{j} + \frac{\pi\bs{j}}{2e^2 T\tau_{vv}\ln 2},
\end{equation}
\begin{eqnarray}
\label{slqeq3dp}
{\cal R}_H\bs{j}\times \bs{e}_{\bs{B}}
= {\cal R}_0 \frac{e}{2T\ln 2}\bs{Q},
\end{eqnarray}
\begin{eqnarray}
\label{slpeq3dp}
{\cal R}_H\bs{j}\times \bs{e}_{\bs{B}}
= e{\cal R}_0 \bs{P} + \frac{\pi\bs{\cal C}}{2e^2 T\tau_{ss}\ln 2}.
\end{eqnarray}
\end{subequations}
The parameters ${\cal R}_0$ and ${\cal R}_H$ are given by
Eq.~(\ref{r0rh}) evaluated at $\mu=0$.

At this point, the essential role of disorder becomes
self-evident. Indeed, in the absence of disorder ${\cal R}_0=0$ and
then Eq.~(\ref{slqeq3dp}) becomes senseless, at least when the system
is subjected to external magnetic field. Physically, this means that
in the absence of disorder our original assumption of the steady state
becomes invalid: under external bias the energy current increases
indefinitely. 

In the absence of magnetic field the electric current is decoupled. In
this case, the electrical resistivity of graphene can be read off
Eq.~(\ref{sljeq3dp}) [using Eqs.~(\ref{kt}) and (\ref{r0rh}) at the
  neutrality point]
\begin{equation}
\label{res-rdp}
R(B=0; \mu=0) = 
\frac{\pi}{2e^2 T\ln 2}\left[\tau^{-1}+\tau_{ee}^{-1}(0)\right].
\end{equation}

If the system is subjected to an external magnetic field, then all
three macroscopic currents are entangled. Using Eqs.~(\ref{aim0}),
(\ref{slqeq3dp}), and (\ref{slpeq3dp}), we find the following
expression for the vector ${\cal K}$ that determines the Lorentz term
in the equation (\ref{sljeq3dp}) for the electric current
\[
\bs{\cal K} = \bs{j}\times\bs{e}_{\bs{B}}\;\kappa{\cal R}_H/[{\cal R}_0\Delta(0)],
\]
where
\[
\kappa = \gamma_0-1
+ \left[\gamma_0-{\cal N}_2(0)\right]
\frac{\Delta(0) -  \gamma_0\tau/\tau_{ss}}
{\Delta(0)-{\cal N}_2(0)\tau/\tau_{ss}}.
\]
Clearly, the direction of the Lorentz term coincides with the
direction of the electric current. Hence, there is no classical Hall
effect at the Dirac point (as expected from symmetry considerations)
\begin{equation}
\label{res-rhdp}
R_{H}(\mu=0) = 0.
\end{equation}

At charge neutrality, carriers from both bands are involved in
scattering processes and the system exhibits nonzero classical
magnetoresistance (similarly to multi-band semiconductors \cite{bsm})
\begin{eqnarray}
&&
R(B;\mu=0) = R(B=0;\mu=0) + \delta R(B;\mu=0),
\nonumber\\
&&
\nonumber\\
&&
\delta R(B;\mu=0) = \frac{{\cal R}_H^2\kappa}{{\cal R}_0\Delta(0)}
\propto \frac{v_g^4\tau}{c^2} \frac{B^2}{T^3}.
\label{res-rbdp}
\end{eqnarray}
The sign of $\delta R(B;\mu=0)$ is determined by the interplay of $\tau$,
$T$, and $\tau_{ss}$. However, using Eqs.~(\ref{d0}) and (\ref{n20}) we find
the coefficient as
\[
\frac{\pi\kappa}{8\ln^32\Delta(0)} \approx
-1.04\; \kappa
\approx 
\frac{1.71+1.03\,\tau/\tau_{ss}}{1+3.59\;\tau/\tau_{ss}}>0.
\]
Thus, our Eq.~(\ref{res-rbdp}) describes {\it positive} magnetoresistance.

Magnetoresistance in graphene was previously calculated within the
two-mode approximation in Ref.~\onlinecite{mu1} where it was found
${\delta}R(B;\mu=0)=[\pi/9\zeta(3)]v_g^4{\tau}c^{-2}B^2/T^3$. This
expression shows the same parameter dependence as our
Eq.~(\ref{res-rbdp}) but with a numerical prefactor
$\pi/9\zeta(3)\approx 0.2904$ which is independent of the interaction
strength. The electron-electron scattering time $\tau_{ss}$ does not
appear in the two-mode approximation. In the ``hydrodynamic'' limit
${\tau\gg\tau_{ss}}$, the prefactor in Eq.~(\ref{res-rbdp}) approaches
the same numerical value as the result of Ref.~\onlinecite{mu1}.

\subsubsection{Transport in mesoscopic samples}

In small enough samples, or in strong enough magnetic fields
$W\ll\ell_R\omega_B^2\tau\tau_{ee}$, boundary conditions become
important and physical quantities become inhomogeneous. The
macroscopic equations acquire gradient terms and have to be considered
alongside the corresponding continuity equations as well as the
Maxwell equations describing the self-consistent electromagnetic
fields. In general, solution to such system of equations is a
formidable computational task that is best approached numerically. The
notable exception is the neutrality point, where the classical Hall
effect is absent (due to exact electron-hole symmetry). In this case,
the electrostatic problem is trivial and we can tackle the problem
analytically. Still, within the three-mode approximation the solution
is rather tedious, see Section~\ref{fse} below. The main qualitative
result is the appearance of the linear magnetoresistance in moderately
strong classical fields for $\ell_R\ll{W}\ll\ell_R\omega_B^2\tau\tau_{ee}$
\[
R \sim B\frac{v_gW}{ecT^2}
\sqrt{\frac{1}{\tau_{ph}}\left[\frac{1}{\tau}+\frac{1}{\tau_{ee}(0)}\right]}.
\]
The result is governed by energy relaxation and quasiparticle
recombination due to electron-phonon interaction. On a qualitative
level, this effect is independent of details of the quasiparticle
spectrum and can also be found in other two-component materials, such
as narrow-band semiconductors, semi-metals, and macroscopically
disordered media at the neutrality point \cite{mrs,gut,mag}.


\section{From kinetic equation to macroscopic description}
\label{hdm}

In this Section we derive the macroscopic equations (\ref{meqs})
describing electronic transport in monolayer graphene in the
interaction-dominated regime.

\subsection{Boltzmann equation approach}

\subsubsection{Kinetic equation}

We begin with the standard (Boltzmann) form of the kinetic equation
\cite{dau,kas,mu1,kin,fos,ryz,mem}:
\begin{equation}
\label{beq0}
\frac{\partial f}{\partial t} + \bs{v}\cdot\frac{\partial f}{\partial \bs{r}}
+\left(e\bs{E} +\frac{e}{c} \;\bs{v}\times\bs{B}\right)\cdot
\frac{\partial f}{\partial \bs{p}} = -\frac{\delta f}{\tau} + {\cal I},
\end{equation}
where $f$ is the distribution function, ${\cal I}$ is the collision
integral due to Coulomb interaction, $\tau$ is the transport impurity
scattering time (which may be energy-dependent), and $\delta f$ is the
non-equilibrium correction 
\begin{equation}
\label{df}
\delta f = f - f^{(0)} = f^{(0)} \left(1-f^{(0)}\right) h = 
-T \frac{\partial f^{(0)}}{\partial \epsilon} h.
\end{equation}
Here $f^{(0)}$ is the equilibrium Fermi-Dirac distribution with the
corresponding temperature $T$. In this paper we consider the
steady-state transport and thus take the distribution function to be
time-independent
\begin{equation}
\label{stst}
\frac{\partial f}{\partial t} = 0.
\end{equation}

\subsubsection{Macroscopic currents}

Let us now introduce macroscopic physical observables. The electric
current is defined as
\begin{subequations}
\label{mc}
\begin{equation}
\label{j0}
\bs{j} = e \sum \bs{v} \delta f,
\end{equation}
where the sum runs over all of the single-particle states. Similarly,
the energy current is defined as
\begin{equation}
\label{q0}
\bs{Q} = \sum \epsilon \bs{v} \delta f.
\end{equation}
Finally we introduce the ``imbalance current'' (cf.
Ref.~\onlinecite{fos})
\begin{equation}
\label{qs}
\bs{P} = \sum {\rm sign}(\epsilon) \bs{v} \delta f.
\end{equation}
\end{subequations}
The appearance of this current reflects the independent conservation
of the number of particles in the upper and lower bands in graphene.

All currents (\ref{mc}) vanish in equilibrium. In the degenerate (or
``Fermi-liquid'') limit, ${\mu\gg{T}}$, the non-equilibrium correction
(\ref{df}) to the distribution function contains a $\delta$-function
\cite{dau}. Thus, the above sums are dominated by the states with
energies close to the chemical potential $\epsilon\sim\mu$ and all
three currents become equivalent, see Eq.~(\ref{flcs}).

\subsubsection{Non-equilibrium distribution function: infinite sample}
\label{ned}

Within the standard linear response theory \cite{dau}, one describes
macroscopic states that are only weakly perturbed from equilibrium by
some external probe. In this case, the non-equilibrium correction
$\delta f$ to the distribution function, or equivalently the function
$h$, see Eq.~(\ref{df}) are linear in the strength of the probe. At
the same time, the function $h$ (which we will hereafter refer to as
the non-equilibrium distribution function) has to be proportional to
the quasiparticle velocity, otherwise the macroscopic currents
(\ref{mc}) will remain zero. Now, within linear response the strength
of the external probe is proportional to the electric current and thus
one can express the non-equilibrium distribution function $h$ as
\[
h = A(\epsilon)\;\bs{j}\cdot\bs{v}.
\]
In an infinite sample all physical quantities are uniform. Moreover,
in the degenerate regime ${A(\epsilon)\rightarrow{A(\mu)}}$. Such a
description is completely equivalent to the standard linear response
theory \cite{dau}, but is more natural in situations where one
passes a current through a sample rather than applies an electric
field, for example in drag measurements \cite{gor,tu1,tu2,meg}.

In nearly neutral graphene, the energy dependence of the distribution
function becomes important. Taking advantage of the collinear
scattering singularity \cite{kas,mu1,kin,mem,meg} we retain only those
terms in the power series of the distribution function $h$ [or the
  prefactor $A(\epsilon)$] in $\epsilon$, which correspond to either
zero modes of the collision integral, or to its eigenmodes with
non-divergent eigenvalues. In general, there are three such terms
\[
A(\epsilon) = A_0 + A_s{\rm sign}(\epsilon) + A_1\epsilon,
\]
where the coefficients $A_i$ can be expressed in terms of the
macroscopic currents by evaluating the sums in Eqs.~(\ref{mc}). The
resulting distribution function allows us to formulate macroscopic or
hydrodynamic equations describing electronic transport in graphene.

If the system is subjected to an external magnetic field, the
direction of the macroscopic currents may deviate from the driving
bias. In this case, we may write the non-equilibrium distribution
function in the form:
\begin{equation}
\label{h0}
h = \frac{2}{e\nu_0Tv_g^2}
\left[C_{\|}(\epsilon)\;\bs{v}\cdot\bs{j} + 
C_\perp(\epsilon) \bs{v}\cdot\left(\bs{j}\times\bs{e}_z\right)\right],
\end{equation}
where $\nu_0$ is the density of states
\begin{equation}
\label{n0}
\nu_0 =
\sum \left( -\frac{\partial f^{(0)}}{\partial \epsilon}\right) = 
\frac{NK}{2\pi v_g^2}, 
\end{equation}
with $N$ being the degeneracy of the single-particle states (in
physical graphene $N=4$).

Based on the above arguments, we truncate the energy-dependent
functions $C_i(\epsilon)$ as follows
\[
C_i(\epsilon) = C_i^{(0)} + C_i^{(s)} {\rm sign}(\epsilon) 
+ C_i^{(1)}\epsilon,
\]
leading to the three-mode approximation for the distribution
function. The coefficients $C_i^{(j)}$ can be found by requiring the
distribution function (\ref{h0}) to yield the physical observables
(\ref{mc}). The resulting expression is somewhat cumbersome and is
given in Appendix~\ref{Ah}. For the subsequent derivation of the
macroscopic equations we only need the energy dependence of the
distribution function for which we use a short-hand notation
(\ref{habc})
\begin{equation}
\label{habc1}
h = 
\frac{2\bs{v}}{e \nu_0 T v_g^2}
\left\{\bs{\cal A} + \bs{\cal B}\frac{\epsilon}{K} 
+ \bs{\cal C}{\rm sign}(\epsilon)\right\}.
\end{equation}
The vectors $\bs{\cal A}$, $\bs{\cal B}$, $\bs{\cal C}$ are given in
Eq.~(\ref{h3m}).

\subsubsection{Macroscopic densities}

The above arguments rely on translational invariance of the infinite
system to establish the fact that all macroscopic physical quantities
are homogeneous. Then the currents can be defined by Eqs.~(\ref{mc}),
while the corresponding densities are determined by the equilibrium
distribution function $f^{(0)}$. As both the currents and densities
are independent of the coordinates and time, the corresponding
continuity equations are trivially satisfied.

Taking into account either sample geometry or local perturbations
leads to non-homogeneous distributions of physical quantities. Within
linear response, the nonuniform deviations of the macroscopic
densities are expected to be small (as determined by the small driving
force) and can be accounted for by an additional term in the
non-equilibrium distribution function similar to Eq.~(\ref{h0}), but
expressed in terms of the densities rather than currents.

To a good approximation, electron and hole numbers in graphene are
conserved independently. Defined as
\begin{subequations}
\label{md}
\begin{equation}
\label{nh}
n_e = \sum\limits_{\epsilon>0} f, \qquad
n_h = \sum\limits_{\epsilon<0} (1-f),
\end{equation}
they can be combined into the total charge density
\begin{equation}
\label{cd}
n = e(n_e-n_h), \quad n = n_0 + \delta n(\bs{r}), \quad
\delta n(\bs{r}) =  e\sum \delta f,
\end{equation}
and the quasiparticle density
\begin{equation}
\label{id}
\rho = n_e+n_h, \, \rho = \rho_0 + \delta \rho(\bs{r}), \,
\delta \rho(\bs{r}) =  \sum {\rm sign}(\epsilon) \delta f.
\end{equation}
Finally, we define the energy density
\begin{eqnarray}
\label{ed}
&&
u = \sum_{\epsilon>0} \epsilon f + \sum\limits_{\epsilon<0} \epsilon  (1-f), 
\\
&&
\nonumber\\
&&
u = u_0 + \delta u(\bs{r}), \quad
\delta u(\bs{r}) =  \sum \epsilon\delta f.
\nonumber
\end{eqnarray}
\end{subequations}
Similarly to Eq.~(\ref{flcs}), all three densities become equivalent
in the degenerate limit
\begin{equation}
\label{flds}
n(\mu\gg T) = \frac{e}{\mu} u(\mu\gg T) = e\rho (\mu\gg T).
\end{equation}

\subsubsection{Non-equilibrium distribution function: mesoscopic sample}
\label{ned-ms}

Consider now a small, mesoscopic sample (still within linear
response). If boundary conditions are important, then the
non-equilibrium distribution function acquires a non-homogeneous term
that can be expressed in terms of the fluctuating densities
(\ref{md}). Now we can write the deviation of the distribution
function (\ref{df}) as follows
\begin{equation}
\label{df-ms}
\delta f = - T \frac{\partial f^{(0)}}{\partial \epsilon}
(h + \delta h),
\end{equation}
where $h$ is given by Eq.~(\ref{habc}) and the extra term $\delta h$
can be written in a similar form
\begin{equation}
\label{dhabc}
\delta h = \frac{1}{e\nu_0 T}
\left[ \mathfrak{a} + \mathfrak{b} \frac{\epsilon}{K} + \mathfrak{c} \; 
{\rm sign}(\epsilon) \right].
\end{equation}
The coefficients $\mathfrak{a}$, $\mathfrak{b}$, and $\mathfrak{c}$
are linear combinations of the inhomogeneous densities (\ref{md}) [cf.
  Eq.~(\ref{dh3m})]. In the degenerate limit
${\mathfrak{a}(\mu\gg{T})=\delta n}$, while
$\mathfrak{b}(\mu\gg{T})=\mathfrak{c}(\mu\gg{T})=0$. At the Dirac
point, these quantities simplify to
\begin{eqnarray}
\label{saim0}
&&
\mathfrak{a}(\mu=0) = \delta n, 
\\
&&
\nonumber\\
&&
\mathfrak{b}(\mu=0) = - \frac{e\delta u}{K\Delta(0)}
+ e\delta\rho\frac{\gamma_0}{\Delta(0)},
\nonumber\\
&&
\nonumber\\
&&
\mathfrak{c}(\mu=0) = \gamma_0 \frac{e\delta u}{K\Delta(0)}
- e\delta\rho\frac{{\cal N}_2(0)}{\Delta(0)}, 
\nonumber
\end{eqnarray}
[cf. Eqs.~(\ref{aim0})].

\subsection{Macroscopic equations: infinite system}

In an infinite system physical quantities are uniform:
\[
\frac{\partial f}{\partial\bs{r}}=0.
\]
Substituting the distribution function (\ref{habc}) into the 
kinetic equation (\ref{beq0}) and integrating over the energies and
momenta of the single-particle states, we find the set of linear 
equations describing the macroscopic currents.

\subsubsection{Electrical current}

Multiplying Eq.~(\ref{beq0}) by $e\bs{v}$ and integrating, we find the
equation for the electric current (\ref{j0})
\begin{subequations}
\label{jeq1}
\begin{equation}
\bs{E} + {\cal R}_H\;\bs{\cal K}\times \bs{e}_{\bs{B}}
= {\cal R}_0 \bs{j} - \frac{\pi}{e^2 K}\bs{\cal I},
\end{equation}
where the coefficients ${\cal R}_0$ and ${\cal R}_H$ are given by
Eq.~(\ref{r0rh}) and the Lorentz-force term contains the vector
\begin{eqnarray}
\label{k}
\bs{\cal K} = e K \sum \bs{v} \frac{\delta f}{\epsilon},
\end{eqnarray}
which for the distribution function (\ref{habc}) has the form (\ref{ka}).

The last term in the right-hand-side of Eq.~(\ref{jeq1}) is the
integrated collision integral 
\begin{equation}
\label{rhs-j}
\bs{\cal I} = e \sum \bs{v} {\cal I} = - \frac{\bs{\cal A}}{\tau_{vv}}
- \frac{\bs{\cal C}}{\tau_{vs}}.
\end{equation}
\end{subequations}
For more details on integration of the collision integral and the
precise expressions for the relaxation rates see
Appendix~\ref{Atauee}. In the two-mode approximation used in
Refs.~\onlinecite{mem,meg} the imbalance current was not introduced
and the rate $\tau_{vs}^{-1}$ did not appear. The rate $\tau_{vv}^{-1}$
was previously introduced in Ref.~\onlinecite{mem}.

In the degenerate limit, the relaxation rates $\tau_{vv}^{-1}$ and
$\tau_{vs}^{-1}$ vanish due to the restored Galilean
invariance. Moreover, the rate $\tau_{vs}^{-1}$ vanishes at the Dirac
point as well
\[
\tau_{vv}^{-1}(\mu\gg T), \tau_s^{-1}(\mu\gg T) \rightarrow 0; \quad 
\tau_{vs}^{-1}(\mu=0) = 0.
\]

Note, that in the general case of energy-dependent impurity scattering
time $\tau(\epsilon)$ the numerical coefficients entering the
equations (\ref{jeq1}) will change. This, however, does not yield any
qualitatively new behavior \cite{us1}. The same applies to all of 
the equations derived below.

\subsubsection{Energy current}

The equation for the energy current can be obtained by multiplying the
kinetic equation (\ref{beq0}) by $\epsilon\bs{v}$ and integrating
similarly to the above. As a result we find
\begin{subequations}
\label{qeq1}
\begin{equation}
{\cal N}_1 \bs{E} + {\cal R}_H \;\bs{j}\times \bs{e}_{\bs{B}}
= {\cal R}_0 \frac{e}{K}\bs{Q} - \frac{\pi}{eK^2}\bs{\cal I}',
\end{equation}
where similarly to Eq.~(\ref{rhs-j}) we define
\begin{equation}
\label{rhs-q}
\bs{\cal I}' = \sum \epsilon \bs{v} {\cal I} = 0.
\end{equation}
\end{subequations}
Physically, the latter equality follows from momentum conservation and
time-reversal properties of the scattering probability. In
double-layer systems, this conclusion applies to the intralayer
collision integral only, see below.

\subsubsection{Imbalance current}

The imbalance current obeys the equation (that can be obtained by
multiplying the kinetic equation by $\bs{v}\,{\rm sign}(\epsilon)$ and
integrating over all single-particle states)
\begin{subequations}
\label{qseq1}
\begin{equation}
\frac{\mu}{K}\bs{E} + {\cal R}_H\;\widetilde{\bs{\cal K}}\times \bs{e}_{\bs{B}}
= e {\cal R}_0 \bs{P} - \frac{\pi}{eK} \bs{\cal I}'',
\end{equation}
where the counterpart of Eq.~(\ref{k}) is [see Eq.~(\ref{kta})]
\begin{equation}
\label{ksa}
\widetilde{\bs{\cal K}} = eK\sum \bs{v} \frac{\delta f}{|\epsilon|}.
\end{equation}
The integrated collision integral in Eq.~(\ref{qseq1}) is given by
\begin{equation}
\label{rhs-p}
\bs{\cal I}'' =  \sum \bs{v} \; {\rm sign}(\epsilon) \; {\cal I}
= - \frac{\bs{\cal A}}{e\tau_{vs}}
- \frac{\bs{\cal C}}{e\tau_{ss}},
\end{equation}
\end{subequations}
see Appendix~\ref{Atauee} for details. In the degenerate limit
\[
\tau_{ss}^{-1}(\mu\gg T) \rightarrow 0.
\]

\subsection{Macroscopic equations in mesoscopic systems}

Is the case of relatively small, mesoscopic samples (see below for
specific conditions) we can no longer rely on translational invariance
and need to determine spatial variations of the physical quantities
from Eq.~(\ref{beq0}). In other words, we need to take into account
the gradient term in the left-hand side of the kinetic equation.

Proceeding similarly to the case of infinite systems, we adopt the
three-mode approximation (\ref{habc}) for the non-equilibrium
distribution function (\ref{df}) and integrate the kinetic equation.
This way, we arrive at the equations (\ref{meqs}), which differ from
the corresponding equations for infinite systems (\ref{jeq1}),
(\ref{qeq1}), and (\ref{qseq1}) by the presence of the gradient
terms in the left-hand side, which originate from integrating the
gradient term $\bs{v}\cdot\bs{\nabla}f$ in Eq.~(\ref{beq0}). This yields
three new macroscopic quantities, which physically describe the flux
density of the electric, energy, and imbalance currents. 

The flux density of the electric current is a tensor that is defined
similarly to the usual momentum flux density \cite{dau} (which can be
called flux density of the mass current)
\begin{equation}
\label{stt}
\Pi_{\alpha\beta} = e \sum_{\epsilon>0} v_\alpha v_\beta f + 
e \sum_{\epsilon<0} v_\alpha v_\beta (1-f) 
= \Pi^{(0)}_{\alpha\beta} + \delta\Pi_{\alpha\beta},
\end{equation}
where $\Pi^{(0)}_{\alpha\beta}$ is the equilibrium tensor, while
$\delta\Pi_{\alpha\beta}$ is the inhomogeneous correction out of
equilibrium.

One of the main steps in the derivation of the usual hydrodynamics
\cite{dau} is to relate higher-rank tensors, such as
$\Pi_{\alpha\beta}$, to the hydrodynamic quantities such as the
macroscopic currents. Depending on the degree of approximation
\cite{dau,cen,gra}, one obtains various expressions for the
higher-rank tensors which lead to various hydrodynamic equations, such
as the Euler or Navier-Stokes equations.

In our linear-response theory the situation is simpler. We already
have the expression for the distribution function in terms of the
macroscopic currents and densities, see Eqs.~(\ref{df-ms}),
(\ref{habc}), and (\ref{dhabc}).  All we need to do is to evaluate the
expression (\ref{stt}) with that distribution function. As a result,
we define the quantity $\Pi$ entering Eq.~(\ref{jeq}):
\[
\delta\Pi_{\alpha\beta} = \delta_{\alpha\beta}\frac{eK}{\pi}\Pi.
\]

Similarly, we define the flux densities of the energy and imbalance
currents. Evaluating the resulting quantities with the distribution
function (\ref{dhabc}) we find 
\begin{subequations}
\label{grs}
\begin{equation}
\label{jgr}
\Pi = \frac{1}{e^2\nu_0} \left[
\mathfrak{a} + \mathfrak{b} {\cal N}_1 + \mathfrak{c} \frac{\mu}{K}
\right],
\end{equation}
\begin{equation}
\label{egr}
\Theta = \frac{1}{e^2\nu_0} \left[
\mathfrak{a}{\cal N}_1 + \mathfrak{b} {\cal N}_2 
+ \mathfrak{c} \frac{T^2}{K^2}\left(\frac{\pi^2}{3}+\frac{\mu^2}{T^2}\right)
\right],
\end{equation}
\begin{equation}
\label{igr}
\Xi = \frac{1}{e^2\nu_0} \left[
\mathfrak{a}\frac{\mu}{K} 
+ \mathfrak{b} \frac{T^2}{K^2}\left(\frac{\pi^2}{3}+\frac{\mu^2}{T^2}\right) 
+ \mathfrak{c}
\right].
\end{equation}
\end{subequations}
The macroscopic equations (\ref{meqs}) are thus derived. Again, all
numerical coefficients are specific to the case of energy-independent
$\tau$.


\section{Finite-size effects in neutral monolayer graphene}
\label{fse}

\subsection{Boundary conditions}

Solutions of the finite-size problems are largely determined by the
boundary conditions. Here we consider the simplest strip geometry: we
assume that our sample has the form of an infinite strip along the
$x$-axis, with the width $W$ in the perpendicular $y$-direction. We
will be interested in the effects of the external magnetic field that
we assume to be directed along the $z$-axis, i.e. perpendicular to the
surface of the sample.

Since the length of the strip is assumed to be very large, all
physical quantities are independent of $x$. Consider the problem,
where a current is being driven through the strip. This fixes the
average current density defined as
\[
\overline{j_x} = \frac{1}{W}\int\limits_{-W/2}^{W/2}dy \; j_x(y).
\]
As there are no contacts along the strip, the $y$-component of any
current must vanish at $y=\pm W/2$:
\begin{subequations}
\begin{equation}
\label{bc}
j_y(\pm  W/2) = Q_y(\pm  W/2) = P_y(\pm  W/2) = 0.
\end{equation}
Combining this argument with the continuity equation (\ref{jceqdp})
yields
\begin{equation}
\label{jy}
j_y = 0.
\end{equation}
\end{subequations}
Finally, charge conservation requires
\[
\int\limits_{-W/2}^{W/2}dy \; \delta n(y)=0.
\]
Our task is to find the average electric field in the strip
\[
\overline{\bs{E}} = \frac{1}{W}\int\limits_{-W/2}^{W/2}dy \; \bs{E}(y),
\]
and hence the sheet resistance of the sample is
\begin{equation}
\label{rdef}
R = \frac{\overline{E_x}}{\overline{j_x}}.
\end{equation}

The electric field satisfies the Maxwell equations (\ref{max}). In
particular, in our geometry it follows from the second of the
equations (\ref{max}) that the $x$-component of the electric field is
a constant
\[
\frac{\partial E_z}{\partial x} - \frac{\partial E_x}{\partial z} = 0
\quad\Rightarrow\quad \frac{\partial E_x}{\partial z} = 0,
\]
\[
\frac{\partial E_y}{\partial x} - \frac{\partial E_x}{\partial y} = 0
\quad\Rightarrow\quad \frac{\partial E_x}{\partial y} = 0,
\]
or in other words
\begin{equation}
\label{ex}
 E_x = \overline{E_x} =const.
\end{equation}

\subsection{Mesoscopic graphene sample at the Dirac point}
\label{slmr}

Consider the set of equations (\ref{meqs}) at the Dirac point. Given
the absence of the Hall effect, the charge density can be assumed to
be uniform. In this case, we find
\begin{subequations}
\label{eqsgdp}
\begin{equation}
\label{jeqsgdp}
\bs{E} + {\cal R}_H\;\bs{\cal K}\times \bs{e}_{\bs{B}}
= {\cal R}_0 \bs{j} + \frac{\pi\bs{j}}{2e^2 T\tau_{vv}\ln 2},
\end{equation}
\begin{equation}
\label{qeqsgdp}
-\frac{\bs{\nabla} \delta u}{2e\nu_0 T\ln 2}  + 
{\cal R}_H\;\bs{j}\times \bs{e}_{\bs{B}}
= {\cal R}_0 \frac{e}{2T\ln 2}\bs{Q},
\end{equation}
\begin{eqnarray}
\label{peqsgdp}
-\frac{\bs{\nabla} \delta \rho}{e\nu_0}  + 
{\cal R}_H\;\bs{j}\times \bs{e}_{\bs{B}}
= e{\cal R}_0 \bs{P} + \frac{\pi\bs{\cal C}}{2e^2 T\tau_{ss}\ln 2},
\end{eqnarray}
where the vectors $\bs{\cal K}$ and $\bs{\cal C}$ [given in
  Eqs.~(\ref{aim0}) above] are
\begin{eqnarray*}
&&
\bs{\cal K} = \frac{\gamma_0-1}{\Delta(0)}\left[\frac{e\bs{Q}}{2T\ln 2} - e\bs{P}\gamma_2\right],
\end{eqnarray*}
and
\begin{eqnarray*}
\bs{\cal C} = \gamma_0 \frac{e\bs{Q}}{2T\Delta(0)\ln 2}
- e\bs{P}\frac{{\cal N}_2(0)}{\Delta(0)},
\end{eqnarray*}
where the numerical coefficients $\gamma_0$ and $\gamma_2$ are given
in Eqs.~(\ref{g0}) and (\ref{g2}). The parameters ${\cal R}_0$ and
${\cal R}_H$ are evaluated at $\mu=0$
\begin{equation}
\label{r0rhdp}
{\cal R}_0 \rightarrow {\cal R}_0(\mu=0)=\frac{\pi}{2e^2T\tau\ln 2},
\end{equation}
\begin{equation}
{\cal R}_H \rightarrow {\cal R}_H(\mu=0)=\frac{\pi v_g^2 B}{4ecT^2\ln^2 2}.
\end{equation}
\end{subequations}
The relaxation times $\tau_{vv}$ and $\tau_{ss}$ are evaluated at the
Dirac point as well.

As we have already mentioned, in these equations all quantities are
independent of the coordinate $x$ along the strip, such that $\delta
u=\delta u(y)$ and $\delta \rho = \delta\rho(y)$. Taking into account
Eq.~(\ref{jy}) we notice, that all the vectors in the left-hand sides
of Eqs.~(\ref{qeqsgdp}) and (\ref{peqsgdp}) are directed along the
$y$-axis. Thus, we find that both the energy current and imbalance
current are orthogonal to the electric current and can be written in
the form
\begin{equation}
\label{qpdp}
\frac{e}{2T\ln 2}\bs{Q} = (0,q), \qquad
e\bs{P} = (0,p).
\end{equation}
Consequently, the vector $\bs{\cal K}$ is also pointing in the $y$
direction. Therefore, the $y$-component of Eq.~(\ref{jeqsgdp}) simply
reads $E_y=0$, as it should be. The $x$-component of Eq.~(\ref{jeqsgdp})
now reads
\begin{equation}
\label{jxdp}
E_x + {\cal R}_H {\cal K}_y = {\cal R}_j j_x, \qquad
{\cal R}_j = {\cal R}_0 + \frac{\pi}{2e^2 T\tau_{vv}\ln 2}.
\end{equation}
The remaining equations (\ref{qeqsgdp}) and (\ref{peqsgdp}), as well
as the corresponding continuity equations (\ref{iceqdp}) and
(\ref{eceqdp}) contain only $y$ components. The continuity equations
can be re-written as follows
\begin{subequations}
\label{ce0}
\begin{equation}
\frac{d}{dy}
\begin{pmatrix}
q \cr p
\end{pmatrix}
= - \widehat{\cal T}_{ph} 
\begin{pmatrix}
\widetilde{\delta u} \cr \widetilde{\delta\rho}
\end{pmatrix},
\end{equation}
where $\widetilde{\delta u} = e\delta u/K$ and
$\widetilde{\delta\rho}=e\delta\rho$, and [cf. Eq.~(\ref{ceqsdp})]
\begin{equation}
\label{m1}
\widehat{\cal T}_{ph}=-\frac{1}{\Delta(0)}
\begin{pmatrix}
\frac{1}{\tau_{Eb}}+\frac{\gamma_0}{\tau_{Ec}} & 
           -\frac{{\cal N}_2(0)}{\tau_{Ec}}-\frac{\gamma_0}{\tau_{Eb}}\cr
-\frac{1}{\tau_{Ib}}-\frac{\gamma_0}{\tau_{Ic}} & 
           \frac{{\cal N}_2(0)}{\tau_{Ic}}+\frac{\gamma_0}{\tau_{Ib}}
\end{pmatrix}.
\end{equation}
\end{subequations}

Combining the continuity equations (\ref{ce0}) with the linear
response equations (\ref{qeqsgdp}) and (\ref{peqsgdp}), we find
\begin{subequations}
\label{eqsdp}
\begin{equation}
\label{qypy}
\frac{1}{e^2\nu_0}
\frac{d^2}{dy^2}
\begin{pmatrix}
q \cr p
\end{pmatrix}
= \widehat{\cal T}_{ph}\widehat{\cal M}_{\cal R}
\begin{pmatrix}
q \cr p
\end{pmatrix}
+
\frac{{\cal R}_H}{{\cal R}_j} E_x\widehat{\cal T}_{ph}
\begin{pmatrix}
1 \cr 1
\end{pmatrix},
\end{equation}
where we have excluded the $y$-dependent electric current using
Eq.~(\ref{jxdp}). The resistance matrix $\widehat{\cal M}_{\cal R}$ is given
by
\begin{equation}
\label{m}
\widehat{\cal M}_{\cal R} =
\begin{pmatrix}
{\cal R}_0 - \delta {\cal R} & \delta {\cal R}\gamma_2 \cr
-{\cal R}_q - \delta {\cal R} & 
{\cal R}_0 + {\cal R}_q\gamma_1+\delta {\cal R}\gamma_2
\end{pmatrix},
\end{equation}
where
\begin{equation}
\label{dr}
\delta {\cal R} = - \frac{{\cal R}_H^2}{{\cal R}_j\Delta(0)}
\left(\gamma_0-1\right),
\end{equation}
\begin{equation}
\label{rq}
{\cal R}_q = - \frac{\gamma_0}{\Delta(0)} \frac{\pi}{2e^2 T\tau_{ss}\ln 2},
\end{equation}
\begin{equation}
\label{rp}
\gamma_1 = \frac{{\cal N}_2(0)}{\gamma_0}\approx 2.3721.
\end{equation}
\end{subequations}
Note, that the same matrix appears in Eqs.~(\ref{sleq3dp}), if one
writes the second and the third equations (\ref{slqeq3dp}) and
(\ref{slpeq3dp}) in matrix form.
 
The differential equation (\ref{qypy}) admits a formal matrix
solution. Using the hard-wall boundary conditions (\ref{bc}) and
averaging over the width of the sample we find
\begin{subequations}
\label{mr1}
\begin{equation}
\begin{pmatrix}
\overline{q} \cr \overline{p}
\end{pmatrix}
= \frac{{\cal R}_H}{{\cal R}_j} E_x
\left[\frac{\tanh\left(\widehat{{\cal L}_{ph}^{-1}}\;W/2\right)}
           {\widehat{{\cal L}_{ph}^{-1}}\;W/2} -1 \right]
\widehat{\cal M}_{\cal R}^{-1}
\begin{pmatrix}
1 \cr 1
\end{pmatrix},
\end{equation}
where
\begin{equation}
\label{lph}
\widehat{{\cal L}_{ph}^{-1}} = \sqrt{e^2\nu_0\widehat{\cal T}_{ph}\widehat{\cal M}_{\cal R}}.
\end{equation}
\end{subequations}

Now, we use the solution (\ref{mr1}) to determine the auxiliary
quantity ${\cal K}_y$ 
\begin{eqnarray}
\label{ky}
&&
\overline{{\cal K}}_y = 
\frac{\gamma_0-1}{\Delta(0)}
\left(\overline{q} - \gamma_2 \overline{p}\right),
\end{eqnarray}
which we then use in Eq.~(\ref{jxdp}) in order to find the resistance
of the sample:
\begin{equation}
\label{mr}
R = \frac{{\cal R}_j}{1-\delta{\cal R} 
\begin{pmatrix}
1 & -\gamma_2
\end{pmatrix}
\left[\frac{\tanh\left(\widehat{{\cal L}_{ph}^{-1}}\;W/2\right)}
           {\widehat{{\cal L}_{ph}^{-1}}\;W/2} -1 \right]
\widehat{\cal M}_{\cal R}^{-1}
\begin{pmatrix}
1 \cr 1
\end{pmatrix}
}.
\end{equation}
This is the final result of this Section. Here the field-dependent
resistance $\delta{\cal R}$ is given in Eqs.~(\ref{dr}), the numerical
coefficient $\gamma_2$ in Eq.~(\ref{g2}), and the matrices
$\widehat{\cal M}$ and $\widehat{{\cal L}_{ph}^{-1}}$ are defined by
Eqs.~(\ref{m}), (\ref{lph}), and (\ref{m1}).

\begin{figure}
\centerline{\includegraphics[width=0.9\columnwidth]{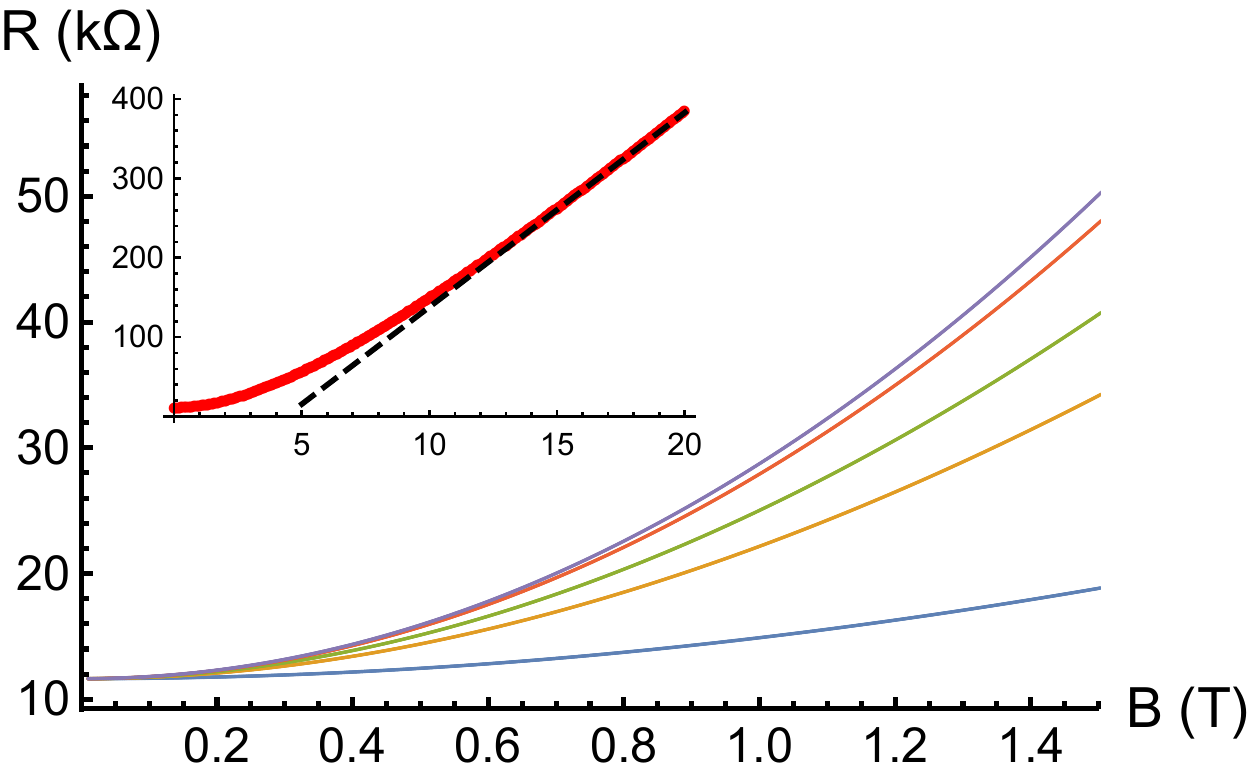}}
\bigskip
\centerline{\includegraphics[width=0.9\columnwidth]{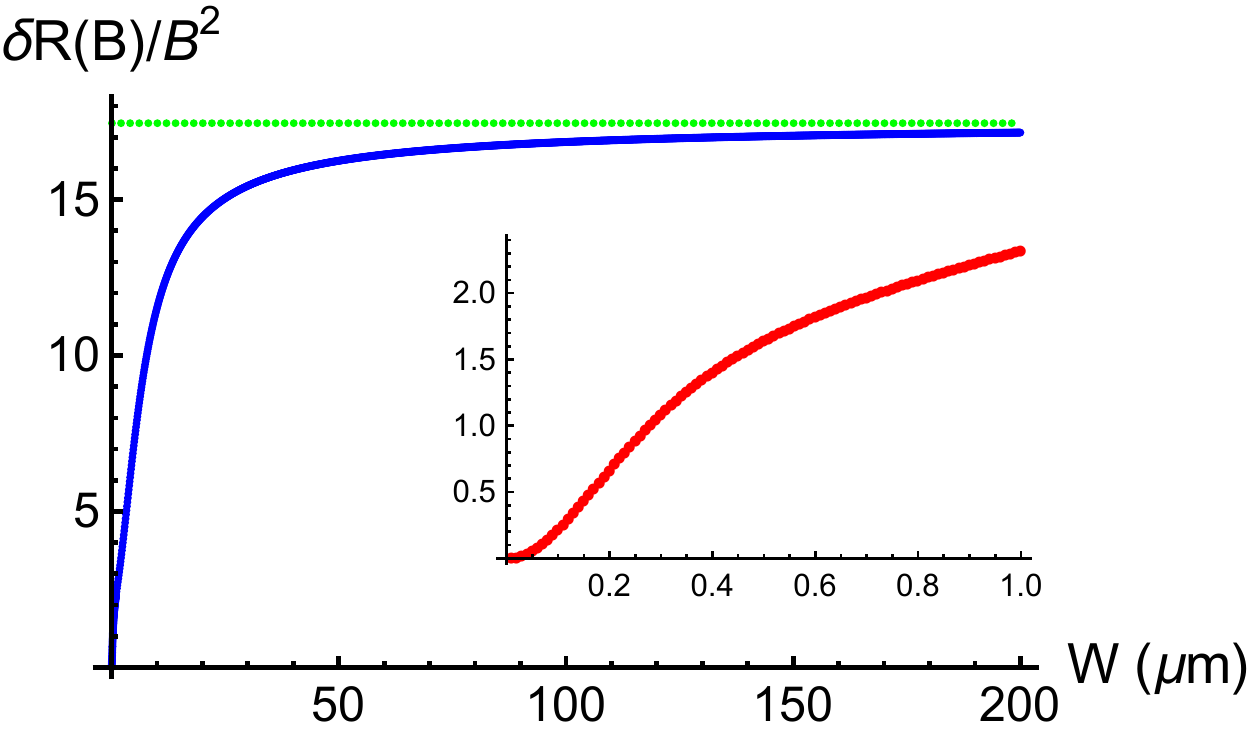}}
\caption{(Color online) Upper panel: magnetoresistance in graphene at
  charge neutrality. The uppermost curve shows the result
  (\ref{res-rbdp}) for a macroscopic sample. The lower curves show the
  result (\ref{mr}) for sample widths $W=100,20,10,2\;\mu$m (top to
  bottom). The results are calculated for realistic values of
  parameters: ${T=240}$\;K, ${\tau^{-1}=50}$\;K,
  ${\tau_{ee}=0.2\tau}$, ${\tau_{ph}=20\tau}$. The inset illustrates
  the linear magnetoresistance for $W=0.1\;\mu$m. The dashed line is a
  guide to the eye. Lower panel: curvature of the above
  magnetoresistance in weak fields (in units of k$\Omega$/T$^2$) as a
  function of $W$. Green line shows the prefactor in
  Eq.~(\ref{res-rbdp}). The inset shows the region ${W<1\;\mu}$m.}
\label{fig:lmr}
\end{figure}

Qualitative behavior of the result (\ref{mr}) is determined by the
interplay of sample geometry, magnetic field, and electron-phonon
scattering.

In the most narrow samples (formally, in the limit
${W\rightarrow{0}}$) the square bracket in Eq.~(\ref{mr}) vanishes and
the resulting resistance is independent of the magnetic field (see
Fig.~\ref{fig:lmr}). Physically, this happens when the electron-phonon
length scale $\ell$ given by the largest eigenvalue of the operator
(\ref{lph}) exceeds the sample width, $W\ll\ell$.

In widest samples, ${W\gg\ell_R\omega_B^2\tau\tau_{ee}}$, [here
  $\ell_R$ is the recombination length given by the smallest
  eigenvalue of the operator (\ref{lph})] the width-dependent term in
Eq.~(\ref{mr}) can be neglected and we reproduce
Eq.~(\ref{res-rbdp}) as
\[
1+\delta{\cal R} 
\begin{pmatrix}
1 & -\gamma_2
\end{pmatrix}
\widehat{\cal M}_{\cal R}^{-1}
\begin{pmatrix}
1 \cr 1
\end{pmatrix}
=\left[1+\frac{{\cal R}_H^2}{{\cal R}_0{\cal R}_j} \frac{\kappa}{\Delta(0)}\right]^{-1}.
\]
The result (\ref{res-rbdp}) is shown by the top curve in
Fig.~\ref{fig:lmr}, where we present magnetoresistance in graphene at
charge neutrality (\ref{mr}) for samples of different widths and for
realistic sample parameters.

In narrower samples the magnetoresistance (\ref{mr}) weakens, see
Fig.~\ref{fig:lmr}. In classically strong fields, ${{\cal R}_H\gg{\cal
    R}_j}$, one finds an intermediate regime,
${\ell_R\ll{W}\ll\ell_R\omega_B^2\tau\tau_{ee}}$, where the system
exhibits linear magnetoresistance
\begin{equation}
\label{lmr}
R \sim  
B \frac{v_g}{c}\sqrt{\frac{{\cal R}_jW^2}{T^3\tau_{ph}}}.
\end{equation}
The recombination length is inversely proportional to the magnetic
field $\ell_R\sim [cT/(ev_gB)]\sqrt{\tau_{ph}/\tau_{ee}}$. Linear
magnetoresistance is illustrated in the inset in Fig.~\ref{fig:lmr}.


\section{Transport properties of double-layer systems}
\label{dlg}

Double-layer systems are often used to study transport properties of
two-dimensional systems. In comparison to single-layer devices, one
can can study two additional phenomena: (i) the relatively weak effect
of the second layer on the single-layer transport properties, and (ii)
the strong Coulomb drag effect. The latter is due to interlayer
electron-electron scattering and is important only in the academic
case of disorder-free graphene in the degenerate limit, where it
provides the only source of resistance. In all other cases, the effect
is relatively small due to the weakness of the interlayer interaction.
On the other hand, the drag effect in double-layer systems
\cite{gor,tu1,tu2,meg} is solely due to the interlayer interaction and
has no counterpart in non-interacting systems. Given the extensive
theoretical literature devoted to Coulomb drag (see
Refs.~\onlinecite{mem,meg,us1,thd,lev} and references therein), here
we focus on the two following issues. Firstly, we compute the leading
correction to the Fermi-liquid prediction for the drag coefficient in
the degenerate regime $\mu\gg{T}$. Secondly, we discuss the drag
effect at charge neutrality, where our theory provides microscopic
justification to the phenomenological treatment of the effect of giant
magnetodrag at charge neutrality given in Ref.~\onlinecite{meg}.

Transport properties of double-layer systems can be described within the 
same macroscopic approach to the Boltzmann equation as we have used above 
in the context of monolayer graphene. Now we introduce the system of
two coupled kinetic equations similar to Eq.~(\ref{beq0}):
\begin{eqnarray}
\label{beq12}
&&
\frac{\partial f_1}{\partial t}
+ \bs{v}_1\cdot\frac{\partial f_1}{\partial\bs{r}} + \left( e\bs{E}_1 
+ \frac{e}{c} \bs{v}_1 \times\bs{B}\right)\cdot\frac{\partial f_1}{\partial\bs{p}}=
\\
&&
\nonumber\\
&&
\qquad\qquad\qquad\qquad
= - \frac{\delta f_1}{\tau} + {\cal I}_{11}(f_1) + {\cal I}_{12}(f_1, f_2),
\nonumber\\
&&
\nonumber\\
&&
\frac{\partial f_2}{\partial t}
+ \bs{v}_2\cdot\frac{\partial f_2}{\partial\bs{r}} + \left( e\bs{E}_2 
+ \frac{e}{c} \bs{v}_2 \times\bs{B}\right)\cdot\frac{\partial f_2}{\partial\bs{p}}=
\nonumber\\
&&
\nonumber\\
&&
\qquad\qquad\qquad\qquad
= - \frac{\delta f_2}{\tau} + {\cal I}_{22}(f_2) + {\cal I}_{21}(f_1, f_2).
\nonumber
\end{eqnarray}
Now the distribution functions $f_i$ carry the layer index
$i=1,2$. The single-layer collision integrals ${\cal I}_{ii}(f_i)$ are
the same as one used in the above discussion of monolayer graphene,
see Eq.~(\ref{Ai0}) and Appendix~\ref{Atauee} for details. The
interlayer coupling is described by the inter-layer collision
integrals ${\cal I}_{12}(f_1, f_2)$ and ${\cal I}_{21}(f_1, f_2)$, see
Appendix~\ref{Ataud}.

\subsection{Infinite system}

Within linear response, deviations of the distribution functions
$f_i$ from equilibrium can be described by Eq.~(\ref{df}). In an
infinite system, we can still use the three-mode approximation 
(\ref{habc}) for the non-equilibrium distribution functions $h_i$
\begin{equation}
\label{habc12}
h_i = 
\frac{2\bs{v}}{e \nu_i T v_g^2}
\left\{\bs{\cal A}_i + \bs{\cal B}_i\frac{\epsilon}{K_i} 
+ \bs{\cal C}_i{\rm sign}(\epsilon)\right\}.
\end{equation}
The vectors in Eq.~(\ref{habc12}) can be read off Eq.~(\ref{h3m}),
with the self-evident addition of the layer index.

\subsubsection{Macroscopic equations}

Here we would like to describe the double-layer system similarly to
the above macroscopic description of monolayer graphene.  Integrating
the kinetic equations (\ref{beq12}) we obtain the following equations
for the macroscopic currents (here $i$ refers to a layer, while $j$
to the other layer)
\begin{subequations}
\label{eqs12}
\begin{equation}
\label{jeq-dls}
\bs{E}_i + {\cal R}_H^{(i)}\;\bs{\cal K}_i\times \bs{e}_{\bs{B}}
= {\cal R}_0^{(i)} \bs{j}_i - \frac{\pi}{e^2 K_i}\bs{\cal I}_{ii}
- \frac{\pi}{e^2 K_i}\bs{\cal I}_{ij},
\end{equation}
\begin{equation}
\label{qeq-dls}
{\cal N}_1^{(i)} \bs{E}_i + {\cal R}_H^{(i)} \;\bs{j}_i\times \bs{e}_{\bs{B}}
= {\cal R}_0^{(i)} \frac{e}{K_i}\bs{Q}_i - \frac{\pi}{eK_i^2}\bs{\cal I}'_{ij},
\end{equation}
\begin{equation}
\label{qseq-dls}
\frac{\mu_i}{K_i}\bs{E}_i 
+ {\cal R}_H^{(i)}\;\widetilde{\bs{\cal K}}_{i}\times \bs{e}_{\bs{B}}
= e {\cal R}_0^{(i)} \bs{P}_i - \frac{\pi}{eK_i} \bs{\cal I}''_{ii}
- \frac{\pi}{eK_i} \bs{\cal I}''_{ij}.
\end{equation}
\end{subequations}
Here the intralayer collision integrals $\bs{\cal I}_{ii}$ and
$\bs{\cal I}''_{ii}$ are still described by Eqs.~(\ref{rhs-j}) and
(\ref{rhs-p}), respectively, with the obvious addition of the layer
index. The interlayer collision integrals are described in detail in
Appendix~\ref{Ataud}. One can recast them in terms of relaxation rates
and rewrite the equations (\ref{eqs12}) in the form (\ref{meqs}).
The resulting equations contain a rather large number of
terms. Therefore, below we will discuss the most interesting limiting
cases, where they can be significantly simplified.

\subsubsection{Coulomb drag in degenerate limit}
\label{cddl}

In the degenerate limit Coulomb drag can be described be means of the
generalized Ohm's (or Drude) equations \cite{mem} with the
phenomenological term describing interlayer friction by means of the
corresponding scattering time $\tau_D$. It is well known \cite{us1},
however, that the traditional Fermi-liquid theory of Coulomb drag is
applicable only for very large densities, far beyond the current
experimental range \cite{gor,tu1,tu2,meg}.

Leading corrections to the Fermi-liquid results can be described in
terms of small deviations of the energy and imbalance currents from
their limiting values (\ref{flcs}). It is intuitively clear that the
imbalance current approaches the limiting value exponentially. In
contrast, the energy current is expected to exhibit power law
corrections. These can be demonstrated by the following arguments.

The drag measurement is performed by passing a current
$\bs{j}_1=j_1\bs{e}_x$ through one of the layers (the active layer)
and measuring the induced electric field (or voltage) in the other,
passive layer. Consider for simplicity identical, macroscopic
layers. In the degenerate regime, we may set $e\bs{P}_1=\bs{j}_1$
(since the deviations from this equality are exponentially small in
$T/\mu$), neglect small differences between various interlayer
relaxation rates, disregard intralayer interaction effects, and assume
interlayer thermalization that yields [see, e.g., Eq.~(\ref{trm})]
\[
\frac{e}{\mu}\bs{Q}_2 = \frac{e}{\mu}\bs{Q}_1 - {\cal N}_1\bs{j}_1.
\]
As a result, the macroscopic equations have the
form
\begin{subequations}
\label{eqs12fl}
\begin{equation}
\label{jeq-dls-fl}
\bs{E}_1 + {\cal R}_H\;\bs{\cal K}_1\times \bs{e}_{\bs{B}}
= \left({\cal R}_0 + {\cal R}_D\right)\bs{j}_1,
\end{equation}
\begin{equation}
\label{qeq-dls-fl}
{\cal N}_1 \bs{E}_1 + \frac{{\cal N}_1+1}{2}{\cal R}_H \;\bs{j}_1\times \bs{e}_{\bs{B}}
= {\cal R}_0 \frac{e}{\mu}\bs{Q}_1 + {\cal N}_1{\cal R}_D\bs{j}_1,
\end{equation}
\begin{equation}
\label{jeq2-dls-fl}
\bs{E}_2 + {\cal R}_H\;\bs{\cal K}_2\times \bs{e}_{\bs{B}}
= - {\cal R}_D\bs{j}_1,
\end{equation}
\end{subequations}
where ${{\cal R}_D=\pi/(e^2\mu\tau_{D})}$ [see Eq.~(\ref{taud})] is
the standard drag resistivity. The auxiliary vectors in the Lorentz
terms read
\[
\bs{\cal K}_1 \approx \bs{j}_1 - \left(\frac{e}{\mu}\bs{Q}_1 - {\cal N}_1\bs{j}_1\right),
\qquad
\bs{\cal K}_2 \approx - \frac{e}{\mu}\bs{Q}_2.
\]

Neglecting small deviations of the energy current in the active layer
from its limiting value $(e/\mu)\bs{Q}_1=\bs{j}_1$, we find the
standard drag effect (defined according to Refs.~\onlinecite{gor,meg})
\begin{equation}
\label{drag-def}
\bs{Q}_2 =0 \quad\Rightarrow\quad R_D=\frac{E_{2x}}{j_1} = - {\cal R}_D,
\end{equation}
which is independent of magnetic field. 

In contrast, taking into account a small deviation of $\bs{Q}_1$
from its limiting value, we find that the leading correction to
$R_D$ depends on magnetic field
\begin{equation}
\label{cdfl}
R_D=-{\cal R}_D + 
\frac{\pi^2T^2}{6\mu^2} 
\frac{{\cal R}_H^2{\cal R}_0}{{\cal R}_0^2+{\cal R}_H^2}.
\end{equation}
Same calculation also yields the Hall drag resistivity:
\begin{equation}
\label{hdfl}
R_{DH}=\frac{E_{2y}}{j_1}=-\frac{\pi^2T^2}{6\mu^2} 
\frac{{\cal R}_H^3}{{\cal R}_0^2+{\cal R}_H^2}.
\end{equation}
In contrast to the traditional theories of Coulomb drag, the above
results contain contributions that do not directly depend on any
interlayer electron-electron scattering rate. Instead, this is the
effect of interlayer thermalization.

At the same time, the presence of the second layer leads to the
appearance of magnetoresistance in the first layer (which vanishes in
the limit $\mu\rightarrow\infty$)
\begin{equation}
\label{mral}
R(\mu\gg T) = {\cal R}_0 + {\cal R}_D + \frac{\pi^2T^2}{6\mu^2} 
\frac{{\cal R}_H^2{\cal R}_0}{{\cal R}_0^2+{\cal R}_H^2},
\end{equation}
as well as a small correction to the Hall coefficient
\begin{equation}
\label{rhal}
R_H(\mu\gg T) = {\cal R}_H - 
\frac{\pi^2T^2}{6\mu^2} 
\frac{{\cal R}_H^3}{{\cal R}_0^2+{\cal R}_H^2}.
\end{equation}

\begin{figure}
\centerline{\includegraphics[width=0.9\columnwidth]{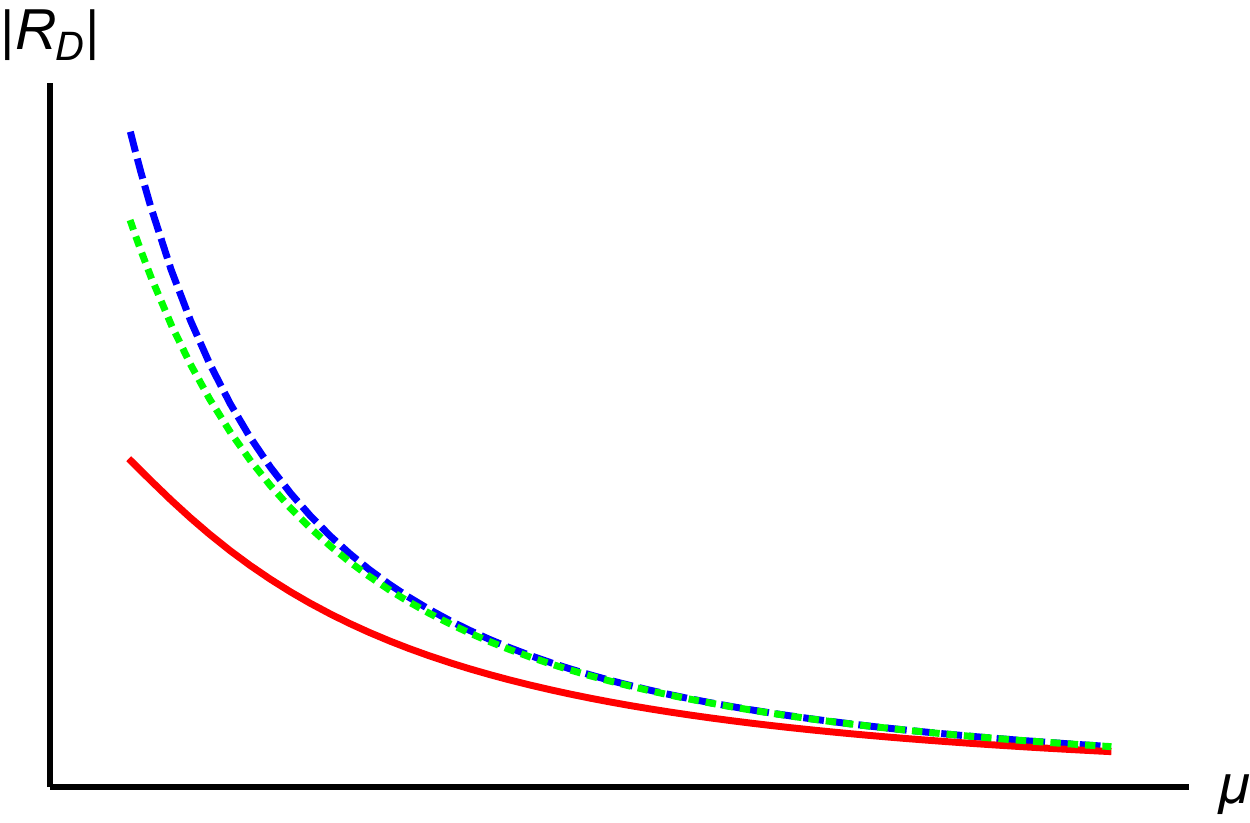}}
\caption{(Color online) Schematic illustration of corrections to the
  Fermi-liquid predictions for the drag coefficient. The blue dashed
  curve represents the Fermi-liquid result ${\cal R}_D\sim
  T^2/\mu^2$. The green dotted curve represents the exponential
  approach to ${\cal R}_D$ within the two-mode approximation that
  retains the electric and imbalance currents. The red solid line
  shows the result (\ref{cdfl}) of the three-mode approximation
  approaching the Fermi-liquid regime as a power law in $T/\mu$.}
\label{fig:rdfl}
\end{figure}

The above corrections exhibit power-law dependence on the small ratio
$T/\mu$. This is in contrast to the exponential approach to the
Fermi-liquid limit that was found within the two-mode approximation in
Ref.~\onlinecite{meg}, as illustrated in Fig.~\ref{fig:rdfl}. Indeed,
the phenomenological model of Ref.~\onlinecite{meg} included the
electric and imbalance currents. The latter approaches its limiting
value $e\bs{P}_1=\bs{j}_1$ only exponentially. Hence the exponentially
small magnetodrag in doped graphene found in Ref.~\onlinecite{meg} (in
notable disagreement with experimental data) is an artifact of
neglecting the energy current in the simplified phenomenological
model. At the same time, at charge neutrality the imbalance current is
proportional to the energy current while both are orthogonal to
$\bs{j}$.  Hence the phenomenological model of Ref.~\onlinecite{meg}
captures qualitative physics at the Dirac point, see below.

\subsubsection{Macroscopic theory at the neutrality point}

At the charge neutrality (or double Dirac) point we may consider the
two layers to be identical. With the help of the thermalization
conditions (\ref{trmdp}) and Eqs.~(\ref{aim0}), we find the following
macroscopic description of an infinite double-layer system
[cf. Eqs.~(\ref{sleq3dp})]. The first layer is described by the
equations [with the auxiliary vectors given by Eqs.~(\ref{aim0})]
\begin{subequations}
\label{eqs12dp}
\begin{equation}
\label{jeq12dp}
\bs{E}_1 + {\cal R}_H\;\bs{\cal K}_1\times \bs{e}_{\bs{B}}
= \left[{\cal R}_j + 
\frac{\pi}{2e^2 T\tau_{vv,12}\ln 2}
\right]\bs{j}_1,
\end{equation}
\begin{equation}
\label{qeq12dp}
\frac{1}{2}{\cal R}_H\;\bs{j}_1\times \bs{e}_{\bs{B}}
= {\cal R}_0^{(1)} \frac{e}{2T\ln 2}\bs{Q}_1,
\end{equation}
\begin{equation}
\label{peq12dp}
\frac{1}{2}{\cal R}_H\;\bs{j}_1\times \bs{e}_{\bs{B}}
= e{\cal R}_0^{(1)} \bs{P}_1 + 
\frac{\pi\left(\tau_{ss}^{-1}+\tau_{ss,12}^{-1}\right)}{2e^2 T\ln 2}\bs{\cal C}_1.
\end{equation}
The energy and imbalance currents in the second layer are determined
by the thermalization conditions (\ref{trmdp}). The equation for the
electric current is similar to Eq.~(\ref{jeq12dp}). However, if we
consider the typical setup for drag measurements, where no electric
current is allowed to flow in the second layer, then the equation
simplifies to
\begin{equation}
\label{jeq21dp}
\bs{E}_2 + {\cal R}_H\;\bs{\cal K}_1\times \bs{e}_{\bs{B}}
=0.
\end{equation}
\end{subequations}

The solution of the equations (\ref{eqs12dp}) is identical to that
described in Section~\ref{sum}. The presence of the second layer does
not significantly change transport in the first layer, in particular,
Eqs.~(\ref{eqs12dp}) still predict positive magnetoresistance. The
Hall classical effect does not appear at charge neutrality as it
should be.

For the second layer, this theory predicts {\it positive} Coulomb drag
[defined in Eq.~(\ref{drag-def})] in agreement with qualitative
arguments of Ref.~\onlinecite{meg}.  In order to explain the
experimentally observed negative drag \cite{gor,meg} the theory needs
to be refined as follows: (i) the finite width $W$ of the system
should be taken into account; the relative parameter is $W/\ell_R$,
where $\ell_R$ is the phonon-induced relaxation length, see
Section~\ref{fse}; (ii) the above interlayer thermalization procedure
should be improved to take into account the finite interlayer
electron-electron relaxation rate. This is outlined in
Section~\ref{s4b}.

\subsection{Mesoscopic system at charge neutrality}
\label{s4b}

In a mesoscopic system, we need to take into account spatial
inhomogeneity of the macroscopic currents and densities. In this case,
the non-equilibrium distribution function acquires the additional
contribution (\ref{dhabc})
\begin{equation}
\label{dhabc12}
\delta h_i = \frac{1}{e\nu_i T}
\left[ \mathfrak{a}_i + \mathfrak{b}_i \frac{\epsilon}{K_i} 
+ \mathfrak{c}_i \; {\rm sign}(\epsilon) \right].
\end{equation}
Similarly to the situation in monolayer graphene, macroscopic
equations in double-layer systems acquire gradient terms. The
resulting equations contain two copies of Eqs.~(\ref{meqs}) where one
has to add interlayer scattering rates from the right-hand side of
Eqs.~(\ref{eqs12}), two copies of continuity equations similar to
Eqs.~(\ref{ceqsdp}) where one has to include additional contributions
due to interlayer electron-electron interaction (see
Appendix~\ref{ad}), and the Maxwell equations (\ref{max}). A general
solution to this system of equations is rather convoluted. Hence here
we limit ourselves to a qualitative discussion.

Of particular interest is the drag effect at charge neutrality, where
the experiment \cite{gor,meg} shows an unusually strong dependence of
$R_D$ on the external magnetic field, i.e. giant magnetodrag. The
problem of Coulomb drag in graphene at charge neutrality was
previously addressed in Refs.~\onlinecite{meg,lev} based on a
two-fluid approach. As shown in Section~\ref{fse} above, the energy
and imbalance currents in the active layer at the neutrality point are
parallel to each other and orthogonal to the driving current
${\bs{Q}_1\|\bs{P}_1\perp\bs{j}_1}$. Excluding one of these currents
from the macroscopic equations one effectively derives a two-fluid
model. Thus our theory provides a microscopic foundation for the
earlier phenomenological models. The key point is that the currents
$\bs{Q}$ and $\bs{P}$ can be transferred between the layers by means
of the interlayer interaction in contrast to the electric current,
whose transfer is forbidden by the exact electron-hole symmetry at the
Dirac point. 

In the limit of infinitely fast interlayer thermalization (discussed
above in Section~\ref{cddl}) the energy and imbalance currents in the
two layers have the same direction leading to positive drag. Taking
into account finiteness of the corresponding relaxation rates
(Appendix~\ref{ad}) refines the theory in analogy with including
viscous terms into standard hydrodynamic theory \cite{dau,ulf}. The
resulting theory contains four differential equations for the energy
and imbalance currents [cf. Eq.~(\ref{qypy}) in the single-layer
  case]. If the sample is wide enough (i.e. if the width of the sample
$W$ is larger than the phonon-induced recombination length), the
energy and imbalance currents in the two layers flow in the same
direction and the system exhibits positive drag as discussed above. On
the contrary, in narrow samples it is the inhomogeneous energy and
imbalance densities in the two layers that coincide, pushing the
currents in the opposite directions and yielding negative drag
\cite{meg,lev}. Similarly to the discussion in Section~\ref{fse}, the
magnetic field dependence of the result is quadratic in weak fields
and linear in classically strong fields.


\section{Summary}
\label{disc}

We have developed a macroscopic (hydrodynamic-like) description of
electronic transport in graphene. Our approach is based on the
``three-mode'' Ansatz for the non-equilibrium distribution function in
graphene. This Ansatz is justified in the interaction-dominated regime
by the collinear scattering singularity in the collision integral.
Under such assumptions, transport properties of graphene can be
described in terms of the three macroscopic currents, $\bs{j}$,
$\bs{P}$, and $\bs{Q}$. In small, mesoscopic samples physical
properties become inhomogeneous and we need to introduce the
inhomogeneous corrections to the corresponding charge, energy, and
imbalance densities. In that case, the complete set of macroscopic
equations includes three equations (\ref{meqs}) for the currents,
which can be viewed as the generalization of the Ohm's law, three
continuity equations, and the Maxwell equations, describing the
self-consistent electromagnetic field. 

Solving the macroscopic equations, one can find temperature, density,
and geometry (i.e. the system size) dependence of transport
coefficients. For general doping this is a formidable computational
task. However, far away from charge neutrality (in the degenerate or
``Fermi-liquid'' regime) all the three currents become equivalent and
the theory reduces to the single-mode equation (\ref{drude-ms}) with
the Drude transport coefficients (\ref{flr}) as it should, given that
no quantum interference processes were taken into account.

Exactly at the Dirac point, the theory simplifies as well and allows
for analytic solutions. We have shown that graphene at charge
neutrality exhibits strong positive magnetoresistance (\ref{mr}).
Specifically, the resistance behaves quadratically in not too strong
fields, Eq.~(\ref{res-rbdp}), and crosses over to the linear
dependence (\ref{lmr}) once the field increases beyond a certain value
determined by the sample width and quasiparticle recombination rate due
to electron-phonon interaction, see Fig.~\ref{fig:lmr}.

Strong positive magnetoresistance in graphene was observed in
Refs.~\onlinecite{fu1,fu2,job,mg3} at charge neutrality. Our results
qualitatively agree with the experimental data. Moreover, our theory
can be generalized to account for macroscopic inhomogeneities that
were discussed as a possible source of magnetoresistance in
Refs.~\onlinecite{fu1,fu2}. Further experimental studies of
magnetoresistance in high-mobility graphene samples (including the
dependence on the sample width) would be of great interest.

In double-layer systems, our theory provides the microscopic
justification of the phenomenological treatment of the giant
magnetodrag problem suggested in Ref.~\onlinecite{meg}. The three-mode
Ansatz allows for more precise quantitative description of the effect.
In particular, we have calculated the leading correction to the
Fermi-liquid prediction for the drag coefficient in doped graphene.
Physically, the resulting magnetodrag (\ref{cdfl}), as well as Hall
drag (\ref{hdfl}) is due to interlayer thermalization. Treating all
three modes on equal footing allows us to remove the artifacts of
two-mode approximations, see Fig.~\ref{fig:rdfl}.

In this paper we have limited ourselves to linear response theory. A
generalization of our approach to nonlinear hydrodynamics in graphene
will be reported in a subsequent publication \cite{ulf}.

\begin{acknowledgments}

We acknowledge helpful conversations with H. Weber, U. Briskot,
A. Levchenko, L. Ponomarenko, P. Alekseev, V. Kachorovsky,
Yu. Vasiliev, and A. Dmitriev. This work was supported by the Dutch
Science Foundation NWO/FOM 13PR3118, the EU Network Grant InterNoM,
DFG-SPP 1459, DFG-SPP 1666, GIF, and the Humboldt Foundation.

\end{acknowledgments}


\appendix

\begin{widetext}

\section{Non-equilibrium distribution function in the three-mode approximation}
\label{Ah}

In this Appendix we give the complete expression for the
non-equilibrium distribution function $h$ in monolayer graphene in
terms of the three macroscopic currents (\ref{mc}) and densities (\ref{md})
\begin{subequations}
\label{h3m}
\begin{eqnarray}
h = 
\frac{2\bs{v}}{e \nu_0 T v_g^2}
\left\{\bs{\cal A} + \bs{\cal B}\frac{\epsilon}{K} 
+ \bs{\cal C}{\rm sign}(\epsilon)\right\},
\end{eqnarray}
\begin{equation}
\label{a}
\bs{\cal A} = \bs{j} + \frac{\frac{e}{K}\bs{Q} - {\cal N}_1 \bs{j}}{\Delta}
\left[{\cal N}_1 - \frac{\mu T^2}{K^3}\left(\frac{\pi^2}{3} + \frac{\mu^2}{T^2}\right)\right]
+ \frac{e\bs{P} -  \frac{\mu}{K} \bs{j}}{\Delta}
\left[{\cal N}_2\frac{\mu}{K}-{\cal N}_1\frac{T^2}{K^2}
           \left(\frac{\pi^2}{3} + \frac{\mu^2}{T^2}\right)\right],
\end{equation}
\begin{equation}
\label{b}
\bs{\cal B} = \frac{\frac{e}{K}\bs{Q} - {\cal N}_1 \bs{j}}{\Delta}\left(\frac{\mu^2}{K^2}-1\right) 
+ \frac{e\bs{P} -  \frac{\mu}{K} \bs{j}}{\Delta}
\left[\frac{T^2}{K^2}\left(\frac{\pi^2}{3} + \frac{\mu^2}{T^2}\right) 
- {\cal N}_1 \frac{\mu}{K}\right]
\end{equation}
\begin{equation}
\label{c}
\bs{\cal C} = \frac{\frac{e}{K}\bs{Q} - {\cal N}_1 \bs{j}}{\Delta}
\left[\frac{T^2}{K^2}\left(\frac{\pi^2}{3} + \frac{\mu^2}{T^2}\right) 
- {\cal N}_1 \frac{\mu}{K}\right]
+ \frac{e\bs{P} -  \frac{\mu}{K} \bs{j}}{\Delta}\left[{\cal N}_1^2-{\cal N}_2\right]
\end{equation}
\end{subequations}

\begin{subequations}
\label{dh3m}
\begin{eqnarray}
&&
\delta h = \frac{1}{e\nu_0 T}
\left[ \mathfrak{a} + \mathfrak{b} \frac{\epsilon}{K} + \mathfrak{c} \; 
{\rm sign}(\epsilon) \right],
\end{eqnarray}
\begin{equation}
\label{ada}
\mathfrak{a} = \delta n + \frac{\frac{e}{K}\delta u - {\cal N}_1 \delta n}{\Delta}
\left[{\cal N}_1 - \frac{\mu T^2}{K^3}\left(\frac{\pi^2}{3} + \frac{\mu^2}{T^2}\right)\right]
+ \frac{e\delta\rho -  \frac{\mu}{K} \delta n}{\Delta}
\left[{\cal N}_2\frac{\mu}{K}-{\cal N}_1\frac{T^2}{K^2}
           \left(\frac{\pi^2}{3} + \frac{\mu^2}{T^2}\right)\right],
\end{equation}
\begin{equation}
\label{bda}
\mathfrak{b} = \frac{\frac{e}{K}\delta u - {\cal N}_1 \delta n}{\Delta}
\left(\frac{\mu^2}{K^2}-1\right) 
+ \frac{e\delta\rho -  \frac{\mu}{K} \delta n}{\Delta}
\left[\frac{T^2}{K^2}\left(\frac{\pi^2}{3} + \frac{\mu^2}{T^2}\right) 
- {\cal N}_1 \frac{\mu}{K}\right]
\end{equation}
\begin{equation}
\label{cda}
\mathfrak{c} = \frac{\frac{e}{K}\delta u - {\cal N}_1 \delta n}{\Delta}
\left[\frac{T^2}{K^2}\left(\frac{\pi^2}{3} + \frac{\mu^2}{T^2}\right) 
- {\cal N}_1 \frac{\mu}{K}\right]
+ \frac{e\delta\rho -  \frac{\mu}{K} \delta n}{\Delta}\left[{\cal N}_1^2-{\cal N}_2\right]
\end{equation}
\end{subequations}
where ${\cal N}_1$ is a dimensionless quantity proportional to the
carrier density in graphene
\begin{subequations}
\label{n}
\begin{eqnarray}
\label{ndef}
\label{nsum}
\label{nd}
n_0 = \int\limits_{-\infty}^\infty d\epsilon \, \nu(\epsilon)
\left[ f^{(0)}(\epsilon; \mu) - f^{(0)}(\epsilon; 0)\right], \qquad
\sum \epsilon \left( -\frac{\partial f^{(0)}}{\partial \epsilon}\right)
= 2 n_0, \qquad
2 n_0 = {\cal N} \nu_0 \mu = {\cal N}_1 \nu_0 K.
\end{eqnarray}
This dimensionless function depends only on the ratio $x=\mu/T$
and has the following asymptotic behavior:
\begin{eqnarray}
\label{nass}
\label{n1ass}
{\cal N}(x) \approx \left\{
\begin{matrix}
2 - \frac{x^2}{6 \ln 2} + \dots , & x\ll 1, \cr\cr
1 + \frac{\pi^2}{3 x^2} + \dots , & x\gg 1,
\end{matrix}
\right., \qquad\qquad
{\cal N}_1(x) \approx \left\{
\begin{matrix}
\frac{x}{\ln 2}\left[1 - \frac{5x^2}{24 \ln 2} + \dots \right], & x\ll 1, \cr\cr
1 + \frac{\pi^2}{3 x^2} + \dots , & x\gg 1.
\end{matrix}
\right.
\end{eqnarray}
\end{subequations}
Similarly, the dimensionless quantity ${\cal N}_2$ represents a similar sum
\begin{eqnarray}
\label{n2}
\label{n2ass}
{\cal N}_2(x) = \frac{1}{\nu_0 K^2} \sum \epsilon^2
\left(-\frac{\partial f^{(0)}}{\partial\epsilon}\right)
\approx 
\left\{
\begin{matrix}
\frac{9\zeta(3)}{8\ln^32}+ 
\frac{3}{4\ln^22}\left(1-\frac{9\zeta(3)}{16\ln^22}\right)x^2, & x\ll 1, \cr\cr
1+\frac{\pi^2}{x^2}, & x \gg 1.
\end{matrix}
\right.,
\end{eqnarray}
and the dimensionless quantity $\Delta$ is
\begin{eqnarray}
\label{del}
\label{dass}
\Delta = \left[\frac{T^2}{K^2}\left(\frac{\pi^2}{3} + \frac{\mu^2}{T^2}\right)
-{\cal N}_1 \frac{\mu}{K}\right]^2+
\left({\cal N}_2-{\cal N}_1^2\right)\left(\frac{\mu^2}{K^2}-1\right)
\approx 
\left\{
\begin{matrix}
-1.13025 + 0.9348 x^2, & x\ll 1, \cr\cr
- \frac{4\pi^2}{3x^3} e^{-x}, & x\gg 1.
\end{matrix}
\right..
\end{eqnarray}

\section{Relaxation rates due to electron-electron interaction}

\subsection{Monolayer graphene}
\label{Atauee}

Within linear response the collision integral in Eq.~(\ref{beq0}) can
be linearized with the help of Eq.~(\ref{df}) as follows
\begin{eqnarray}
{\cal I} = \sum_{1, 1', 2'} W_{12, 1'2'}  f^{(0)}_1 f^{(0)}_2 \left[1-f^{(0)}_{1'}\right]
\left[1-f^{(0)}_{2'}\right]
\left[ h_{1'} + h_{2'} - h_1 - h_2 \right].
\label{Ai0}
\end{eqnarray}

\subsubsection{Collision term in the equation for the electric current}

Following the usual steps involving introduction of transferred energy
$\omega$ and momentum $\bs{q}$, we find for the integrated collision
integral Eq.~(\ref{rhs-j}) appearing in the equation (\ref{jeq1}) for
the electric current:
\begin{eqnarray}
\label{ij1}
&&
\bs{\cal I} = - \frac{e}{32} \int \frac{d^2q d\omega}{(2\pi)^3}
\frac{|U(\bs{q},\omega)|^2}{\sinh^2(\omega/2T)}
\\
&&
\nonumber\\
&&
\qquad\qquad\qquad
\times
\sum_{1, 1'} 
(2\pi)^3 |\lambda_{\bs{v}\bs{v}'}|^2
\delta(\epsilon_1 - \epsilon_1' + \omega)
\delta(\bs{p}_1 - \bs{p}_1' + \bs{q})
\left[\tanh\frac{\epsilon_1-\mu}{2T} - \tanh\frac{\epsilon_1+\omega-\mu}{2T}\right]
\nonumber\\
&&
\nonumber\\
&&
\qquad\qquad\qquad
\times
\sum_{2,2'} (\bs{v}_2-\bs{v}_2')
(2\pi)^3 |\lambda_{\bs{v}\bs{v}'}|^2
\delta(\epsilon_2 - \epsilon_2' - \omega)
\delta(\bs{p}_2 - \bs{p}_2' - \bs{q})
\left[\tanh\frac{\epsilon_2-\mu}{2T} - \tanh\frac{\epsilon_2-\omega-\mu}{2T}\right]
\nonumber\\
&&
\nonumber\\
&&
\qquad\qquad\qquad
\times
\left[ h_{1'} + h_{2'} - h_1 - h_2 \right].
\nonumber
\end{eqnarray}
Here $|\lambda_{\bs{v}\bs{v}'}|^2$ are the ``Dirac factors''. 

Taking into account the explicit form of the distribution function
(\ref{habc}), summations over states $1, 1'$ and $2, 2'$
factorize. Consequently, one can evaluate them separately. The
resulting expressions can be denoted as follows:
\begin{subequations}
\label{yd}
\begin{eqnarray}
(2\pi)^3 \sum_{1, 1'} 
|\lambda_{\bs{v}\bs{v}'}|^2
\delta(\epsilon_1 - \epsilon_1' + \omega)
\delta(\bs{p}_1 - \bs{p}_1' + \bs{q})
\left[\tanh\frac{\epsilon_1-\mu}{2T} - \tanh\frac{\epsilon_1+\omega-\mu}{2T}\right]
= Y_{00}(\bs{q}, \omega),
\label{y0d}
\end{eqnarray}
\begin{eqnarray}
(2\pi)^3 \sum_{1, 1'} \left(\bs{v}_1'-\bs{v}_1\right)
|\lambda_{\bs{v}\bs{v}'}|^2
\delta(\epsilon_1 - \epsilon_1' + \omega)
\delta(\bs{p}_1 - \bs{p}_1' + \bs{q})
\left[\tanh\frac{\epsilon_1-\mu}{2T} - \tanh\frac{\epsilon_1+\omega-\mu}{2T}\right]
= \frac{\bs{q}}{q} v_g  Y_{0A}(\bs{q}, \omega),
\label{y1d}
\end{eqnarray}
\begin{equation}
(2\pi)^3 \sum_{1, 1'} \left(\bs{v}_1'{\rm sgn}\;\epsilon_1'-\bs{v}_1{\rm sgn}\;\epsilon_1\right)
|\lambda_{\bs{v}\bs{v}'}|^2
\delta(\epsilon_1 - \epsilon_1' + \omega)
\delta(\bs{p}_1 - \bs{p}_1' + \bs{q})
\left[\tanh\frac{\epsilon_1-\mu}{2T} - \tanh\frac{\epsilon_1+\omega-\mu}{2T}\right]
= \frac{\bs{q}}{q} v_g  Y_{0C}(\bs{q}, \omega),
\label{y1sd}
\end{equation}
\begin{eqnarray}
&&
(2\pi)^3\sum_{1, 1'} \left(v_1'^\alpha-v_1^\alpha\right)\left(v_1'^\beta -v_1^\beta \right)
|\lambda_{\bs{v}\bs{v}'}|^2
\delta(\epsilon_1 - \epsilon_1' + \omega)
\delta(\bs{p}_1 - \bs{p}_1' + \bs{q})
\nonumber\\
&&
\nonumber\\
&&
\qquad\qquad\qquad
\qquad\qquad\qquad\qquad
\times
\left[\tanh\frac{\epsilon_1-\mu}{2T} - \tanh\frac{\epsilon_1+\omega-\mu}{2T}\right]
= \frac{q^\alpha q^\beta}{q^2} v_g^2  Y_{AA}(\bs{q}, \omega),
\label{y2d}
\end{eqnarray}
\begin{eqnarray}
&&
(2\pi)^3\sum_{1, 1'} \left(v_1'^\alpha-v_1^\alpha\right)
\left(v_1'^\beta{\rm sgn}\;\epsilon_1' -v_1^\beta{\rm sgn}\;\epsilon_1 \right)
|\lambda_{\bs{v}\bs{v}'}|^2
\delta(\epsilon_1 - \epsilon_1' + \omega)
\delta(\bs{p}_1 - \bs{p}_1' + \bs{q})
\nonumber\\
&&
\nonumber\\
&&
\qquad\qquad\qquad\qquad\qquad\qquad\qquad
\times
\left[\tanh\frac{\epsilon_1-\mu}{2T} - \tanh\frac{\epsilon_1+\omega-\mu}{2T}\right]
= \frac{q^\alpha q^\beta}{q^2} v_g^2  Y_{AC}(\bs{q}, \omega),
\label{y2sd}
\end{eqnarray}
\end{subequations}
All of thus defined functions $Y_{ij}(\bs{q}, \omega)$ obey the trivial
symmetry property
\begin{eqnarray}
\label{ysym}
Y_{ij}(\bs{q}, \omega) = - Y_{ij}(-\bs{q}, -\omega).
\end{eqnarray}

Since the collision integral ${\cal I}$ has the dimension of inverse
time, it is convenient to introduce the transport scattering times due
to Coulomb interaction. Given the multitude of terms in the kinetic
equation, we choose to define several interaction-related time scales.
In the current equation, two such time scales appear (if the arguments
of $Y_i(\bs{q}, \omega)$ have their standard form we omit them for
brevity):
\begin{eqnarray}
\label{tauee}
\frac{1}{\tau_{vv}} = 
\frac{1}{\nu_0} \sqint 
\left( Y_{00} Y_{AA} - Y_{0A}^2 \right),
\qquad {\rm where} \qquad \sqint \dots=\frac{e}{32T}\int\frac{d^2q d\omega}{(2\pi)^3}
\frac{|U(\bs{q},\omega)|^2}{\sinh^2(\omega/2T)} \dots,
\end{eqnarray}
\begin{eqnarray}
\label{taus}
\frac{1}{\tau_{vs}} = 
\frac{1}{\nu_0} \sqint
\left( Y_{00} Y_{AC} - Y_{0A}Y_{0C} \right).
\end{eqnarray}
Both time scales $\tau_{vv}^{-1}$ and $\tau_{vs}^{-1}$ vanish in the
Fermi-liquid limit (physically, due to the restored Galilean
invariance). On the other hand, at charge neutrality $\tau_{vs}^{-1}=0$, since
\begin{equation}
\label{ydp}
Y_{0A}(\mu=0)=Y_{AC}(\mu=0)=0,
\end{equation}
while $\tau_{vv}^{-1}$ remains finite.

Using the above relaxation rates, we can write the integrated collision
integral in equation (\ref{jeq1}) in the form (\ref{rhs-j}).

\subsubsection{Collision term in the equation for the imbalance current}

Treating the collision integral in Eq.~(\ref{qseq1}) in the same way
as Eq.~(\ref{ij1}) above, we find:
\begin{eqnarray}
\label{iqs1}
&&
\bs{\cal I}'' = - \frac{1}{32} \int \frac{d^2q d\omega}{(2\pi)^3}
\frac{|U(\bs{q},\omega)|^2}{\sinh^2(\omega/2T)}
\\
&&
\nonumber\\
&&
\qquad\qquad
\times
\sum_{1, 1'} 
(2\pi)^3 |\lambda_{\bs{v}\bs{v}'}|^2
\delta(\epsilon_1 - \epsilon_1' + \omega)
\delta(\bs{p}_1 - \bs{p}_1' + \bs{q})
\left[\tanh\frac{\epsilon_1-\mu}{2T} - \tanh\frac{\epsilon_1+\omega-\mu}{2T}\right]
\nonumber\\
&&
\nonumber\\
&&
\qquad\qquad
\times
\sum_{2,2'} (\bs{v}_2{\rm sgn}\;\epsilon_2-\bs{v}_2'{\rm sgn}\;\epsilon_2')
(2\pi)^3 |\lambda_{\bs{v}\bs{v}'}|^2
\delta(\epsilon_2 - \epsilon_2' - \omega)
\delta(\bs{p}_2 - \bs{p}_2' - \bs{q})
\left[\tanh\frac{\epsilon_2-\mu}{2T} - \tanh\frac{\epsilon_2-\omega-\mu}{2T}\right]
\nonumber\\
&&
\nonumber\\
&&
\qquad\qquad
\times
\left[ h_{1'} + h_{2'} - h_1 - h_2 \right].
\nonumber
\end{eqnarray}
Following the same line of argument as in the previous Appendix, we
introduce another time scale
\begin{eqnarray}
\label{tauss}
\frac{1}{\tau_{ss}} = 
\frac{1}{\nu_0} \sqint 
\left( Y_{00} Y_{CC} - Y_{0C}^2 \right),
\end{eqnarray}
where we had to introduce another quantity $Y_{CC}$ similarly to Eqs.~(\ref{yd}):
\begin{eqnarray}
&&
(2\pi)^3\sum_{1, 1'} 
\left(v_1'^\alpha{\rm sgn}\;\epsilon_1'-v_1^\alpha{\rm sgn}\;\epsilon_1\right)
\left(v_1'^\beta{\rm sgn}\;\epsilon_1' -v_1^\beta{\rm sgn}\;\epsilon_1 \right)
|\lambda_{\bs{v}\bs{v}'}|^2
\nonumber\\
&&
\nonumber\\
&&
\qquad\qquad\qquad
\times
\delta(\epsilon_1 - \epsilon_1' + \omega)
\delta(\bs{p}_1 - \bs{p}_1' + \bs{q})
\left[\tanh\frac{\epsilon_1-\mu}{2T} - \tanh\frac{\epsilon_1+\omega-\mu}{2T}\right]
= \frac{q^\alpha q^\beta}{q^2} v_g^2  Y_{CC}(\bs{q}, \omega).
\label{y2ssd}
\end{eqnarray}
As a result, the integrated collision term (\ref{iqs1}) takes the form (\ref{rhs-p}).

\subsection{Double-layer system}
\label{Ataud}

\subsubsection{Collision term in the equation for the electric current}

The integrated inter-layer collision integral has a form, similar to Eq.~(\ref{ij1}),
\begin{eqnarray}
\label{ij12}
&&
\bs{\cal I}_{12} = - \frac{e}{32} \int \frac{d^2q d\omega}{(2\pi)^3}
\frac{|U_{12}(\bs{q},\omega)|^2}{\sinh^2(\omega/2T)}
\\
&&
\nonumber\\
&&
\qquad\qquad\qquad
\times
\sum_{1, 1'} 
(2\pi)^3 |\lambda_{\bs{v}\bs{v}'}|^2
\delta(\epsilon_1 - \epsilon_1' + \omega)
\delta(\bs{p}_1 - \bs{p}_1' + \bs{q})
\left[\tanh\frac{\epsilon_1-\mu_2}{2T} - \tanh\frac{\epsilon_1+\omega-\mu_2}{2T}\right]
\nonumber\\
&&
\nonumber\\
&&
\qquad\qquad\qquad
\times
\sum_{2,2'} (\bs{v}_2-\bs{v}_2')
(2\pi)^3 |\lambda_{\bs{v}\bs{v}'}|^2
\delta(\epsilon_2 - \epsilon_2' - \omega)
\delta(\bs{p}_2 - \bs{p}_2' - \bs{q})
\left[\tanh\frac{\epsilon_2-\mu_1}{2T} - \tanh\frac{\epsilon_2-\omega-\mu_1}{2T}\right]
\nonumber\\
&&
\nonumber\\
&&
\qquad\qquad\qquad
\times
\left[ h_{2,1'} + h_{1,2'} - h_{2,1} - h_{1,2} \right],
\nonumber
\end{eqnarray}
except than now the chemical potentials and the non-equilibrium
distribution functions carry the layer index (i.e. $h_{2,1}$ stands
for the distribution function in layer $2$ describing the state $1$)
and the potential $U_{12}(\bs{q},\omega)$ describes interlayer interaction.

Consequently, the auxiliary functions (\ref{yd}) as well as the
densities of states, will now also acquire the layer index. This leads
to a larger number of decay rates in comparison to $\tau_{ee}^{-1}$
and $\tau_{s}^{-1}$. Since most of them vanish at the Dirac point, we express
the collision integral (\ref{ij12}) as follows:
\begin{eqnarray}
\label{ij12a}
&&
\bs{\cal I}_{12} = - \bs{\cal A}_1 \frac{e}{\nu_{01}} \oint Y_{AA}^{(1)}Y_{00}^{(2)}
+ \bs{\cal A}_2 \frac{e}{\nu_{02}} \oint Y_{0A}^{(1)}Y_{0A}^{(2)}
- \bs{\cal B}_1 \frac{ev_g}{\nu_{01}K_1} \oint q Y_{0A}^{(1)}Y_{00}^{(2)}
\\
&&
\nonumber\\
&&
\qquad\qquad\qquad
+ \bs{\cal B}_2 \frac{ev_g}{\nu_{02}K_2} \oint q Y_{0A}^{(1)}Y_{00}^{(2)}
- \bs{\cal C}_1 \frac{e}{\nu_{01}} \oint Y_{AC}^{(1)}Y_{00}^{(2)}
+ \bs{\cal C}_2 \frac{e}{\nu_{02}} \oint Y_{0A}^{(1)}Y_{0C}^{(2)},
\nonumber
\end{eqnarray}
where
\begin{equation}
\oint \dots=\frac{1}{32T}\int\frac{d^2q d\omega}{(2\pi)^3}
\frac{|U_{12}(\bs{q},\omega)|^2}{\sinh^2(\omega/2T)} \dots .
\end{equation}
The first two terms are familiar from the traditional theory of
Coulomb drag \cite{us1}. In particular, the usual ``drag rate''
$\tau_D^{-1}$ is given by the second term
\begin{equation}
\label{taud}
\frac{1}{\tau_D} = \frac{e}{\nu_{02}} \oint Y_{0A}^{(1)}Y_{0A}^{(2)}.
\end{equation}
In the degenerate regime, the relaxation rates in the first two terms
become identical. The traditional theory is then recovered by taking
into account interlayer thermalization, see below.
 
At the neutrality point this expression simplifies significantly. Indeed,
taking into account Eq.~(\ref{ydp}) we find
\begin{equation}
\label{ij12dp}
\bs{\cal I}_{12}(\mu_1=\mu_2=0) = - \bs{\cal A}_1 \frac{e}{\nu_{0}} \oint Y_{AA}Y_{00} 
= - \frac{\bs{\cal A}_1}{\tau_{vv,12}},
\end{equation}
where the layer indices can be omitted since at the neutrality point
the layers are identical to each other. On the other hand, the new
relaxation rate $1/\tau_{ee,12}$ differs from Eq.~(\ref{tauee})
insofar it reflects the interlayer interaction potential
$U_{12}(\bs{q},\omega)$.

\subsubsection{Collision term in the equation for the energy current}

The equation for the energy current is obtained by multiplying the
kinetic equation by $\epsilon\bs{v}$ and integrating over all
states. Then, similarly to Eq.~(\ref{ij12}) we find
\begin{eqnarray}
\label{iq12}
&&
\bs{\cal I}'_{12} = - \frac{1}{32} \int \frac{d^2q d\omega}{(2\pi)^3}
\frac{|U_{12}(\bs{q},\omega)|^2}{\sinh^2(\omega/2T)}
\\
&&
\nonumber\\
&&
\qquad\qquad\qquad
\times
\sum_{1, 1'} 
(2\pi)^3 |\lambda_{\bs{v}\bs{v}'}|^2
\delta(\epsilon_1 - \epsilon_1' + \omega)
\delta(\bs{p}_1 - \bs{p}_1' + \bs{q})
\left[\tanh\frac{\epsilon_1-\mu_2}{2T} - \tanh\frac{\epsilon_1+\omega-\mu_2}{2T}\right]
\nonumber\\
&&
\nonumber\\
&&
\qquad\qquad\qquad
\times
\sum_{2,2'} (\epsilon_2\bs{v}_2-\epsilon_2'\bs{v}_2')
(2\pi)^3 |\lambda_{\bs{v}\bs{v}'}|^2
\delta(\epsilon_2 - \epsilon_2' - \omega)
\delta(\bs{p}_2 - \bs{p}_2' - \bs{q})
\left[\tanh\frac{\epsilon_2-\mu_1}{2T} - \tanh\frac{\epsilon_2-\omega-\mu_1}{2T}\right]
\nonumber\\
&&
\nonumber\\
&&
\qquad\qquad\qquad
\times
\left[ h_{2,1'} + h_{1,2'} - h_{2,1} - h_{1,2} \right],
\nonumber
\end{eqnarray}
where (due to momentum conservation)
\begin{equation}
\epsilon_2\bs{v}_2-\epsilon_2'\bs{v}_2' = v_g^2\bs{q}.
\end{equation}
In contrast to monolayer graphene [see Eq.~(\ref{rhs-q})], the
integrated collision integral in the double-layer system does not
vanish. Similarly to Eq.~(\ref{ij12a}) we find
\begin{eqnarray}
\label{iq12a}
&&
\bs{\cal I}'_{12} = - \bs{\cal A}_1 \frac{v_g}{\nu_{01}} \oint q Y_{0A}^{(1)}Y_{00}^{(2)}
+ \bs{\cal A}_2 \frac{v_g}{\nu_{02}} \oint q Y_{00}^{(1)}Y_{0A}^{(2)}
- \bs{\cal B}_1 \frac{v_g^2}{\nu_{01}K_1} \oint q^2 Y_{00}^{(1)}Y_{00}^{(2)}
\\
&&
\nonumber\\
&&
\qquad\qquad\qquad
+ \bs{\cal B}_2 \frac{v_g^2}{\nu_{02}K_2} \oint q^2 Y_{00}^{(1)}Y_{00}^{(2)}
- \bs{\cal C}_1 \frac{v_g}{\nu_{01}} \oint q Y_{0C}^{(1)}Y_{00}^{(2)}
+ \bs{\cal C}_2 \frac{v_g}{\nu_{02}} \oint q Y_{00}^{(1)}Y_{0C}^{(2)},
\nonumber
\end{eqnarray}
At the neutrality point, the first two terms vanish [similarly to
  Eq.~(\ref{ij12dp})]
\begin{equation}
\label{iq12dp}
\bs{\cal I}'_{12}(\mu_1=\mu_2=0) = - \left(\bs{\cal B}_1 - \bs{\cal B}_2 \right)
 \frac{v_g^2}{\nu_{0}K(0)} \oint q^2 Y_{00}Y_{00}
- \left(\bs{\cal C}_1 - \bs{\cal C}_2 \right)
 \frac{v_g}{\nu_{0}} \oint q Y_{0C}Y_{00}.
\end{equation}

\subsubsection{Collision term in the equation for the imbalance current}

The integrated interlayer collision integral in the equation for the
imbalance current takes the form
\begin{eqnarray}
\label{ip12}
&&
\bs{\cal I}''_{12} = - \frac{1}{32} \int \frac{d^2q d\omega}{(2\pi)^3}
\frac{|U_{12}(\bs{q},\omega)|^2}{\sinh^2(\omega/2T)}
\\
&&
\nonumber\\
&&
\qquad\qquad\qquad
\times
\sum_{1, 1'} 
(2\pi)^3 |\lambda_{\bs{v}\bs{v}'}|^2
\delta(\epsilon_1 - \epsilon_1' + \omega)
\delta(\bs{p}_1 - \bs{p}_1' + \bs{q})
\left[\tanh\frac{\epsilon_1-\mu_2}{2T} - \tanh\frac{\epsilon_1+\omega-\mu_2}{2T}\right]
\nonumber\\
&&
\nonumber\\
&&
\qquad
\times
\sum_{2,2'} \left[\bs{v}_2{\rm sign}(\epsilon_2) -\bs{v}_2'{\rm sign}(\epsilon_2')\right]
(2\pi)^3 |\lambda_{\bs{v}\bs{v}'}|^2
\delta(\epsilon_2 - \epsilon_2' - \omega)
\delta(\bs{p}_2 - \bs{p}_2' - \bs{q})
\left[\tanh\frac{\epsilon_2-\mu_1}{2T} - \tanh\frac{\epsilon_2-\omega-\mu_1}{2T}\right]
\nonumber\\
&&
\nonumber\\
&&
\qquad\qquad\qquad
\times
\left[ h_{2,1'} + h_{1,2'} - h_{2,1} - h_{1,2} \right].
\nonumber
\end{eqnarray}
Similarly to Eqs.~(\ref{ij12a}) and (\ref{iq12a}), we can re-write
Eq.~(\ref{ip12}) as follows
\begin{eqnarray}
\label{ip12a}
&&
\bs{\cal I}''_{12} = - \bs{\cal A}_1 \frac{1}{\nu_{01}} \oint Y_{AC}^{(1)}Y_{00}^{(2)}
+ \bs{\cal A}_2 \frac{1}{\nu_{02}} \oint Y_{0C}^{(1)}Y_{0A}^{(2)}
- \bs{\cal B}_1 \frac{v_g}{\nu_{01}K_1} \oint q Y_{0C}^{(1)}Y_{00}^{(2)}
\\
&&
\nonumber\\
&&
\qquad\qquad\qquad
+ \bs{\cal B}_2 \frac{v_g}{\nu_{02}K_2} \oint q Y_{0C}^{(1)}Y_{00}^{(2)}
- \bs{\cal C}_1 \frac{1}{\nu_{01}} \oint Y_{CC}^{(1)}Y_{00}^{(2)}
+ \bs{\cal C}_2 \frac{1}{\nu_{02}} \oint Y_{0C}^{(1)}Y_{0C}^{(2)}.
\nonumber
\end{eqnarray}
At the neutrality point the above expression simplifies and takes the
form
\begin{equation}
\label{ip12dp}
\bs{\cal I}''_{12}(\mu_1=\mu_2=0) = - \left(\bs{\cal B}_1 - \bs{\cal B}_2 \right)
 \frac{v_g}{\nu_{0}K(0)} \oint q Y_{0C}Y_{00}
- \bs{\cal C}_1 \frac{1}{\nu_{0}} \oint Y_{CC}Y_{00}
+ \bs{\cal C}_2 \frac{1}{\nu_{0}} \oint Y_{0C}Y_{0C}.
\end{equation}

\subsubsection{Interlayer thermalization}
\label{it}

The integrated collision integrals (\ref{iq12a}) and (\ref{ip12a}) 
contain formally diverging expressions
\[
\oint q^2 Y_{00}^{(1)}Y_{00}^{(2)}, \qquad
\oint q Y_{0C}^{(i)}Y_{00}^{(j)}, \qquad
\oint Y_{CC}^{(i)}Y_{00}^{(j)}, \qquad
\oint Y_{0C}^{(1)}Y_{0C}^{(2)}.
\]
The divergence stems from the fact that each of the functions $Y_{00}^{(i)}$, $Y_{0C}^{(i)}$,
and $Y_{CC}^{(i)}$ diverge as $|\omega|\rightarrow v_gq$
\[
Y_{00}^{(i)}\propto \frac{1}{\sqrt{\left|\omega^2-v_g^2q^2\right|}}, \qquad
Y_{0C}^{(i)}(|\omega|>v_gq), Y_{CC}^{(i)}(|\omega|>v_gq)\propto \frac{1}{\sqrt{\omega^2-v_g^2q^2}}.
\]
The diverging part can be separated with the help of the following 
relations
\begin{equation}
\label{y0c}
Y_{0C}^{(i)}(|\omega|>v_gq) = 2 \frac{|\omega|}{v_gq}Y_{00}^{(i)}(|\omega|>v_gq) 
+ \widetilde{Y}_{0C}^{(i)}(|\omega|>v_gq),
\end{equation}
\begin{equation}
\label{ycc}
Y_{CC}^{(i)}(|\omega|>v_gq) = 4 \frac{\omega^2}{v_g^2q^2}Y_{00}^{(i)}(|\omega|>v_gq) 
+ \widetilde{Y}_{CC}^{(i)}(|\omega|>v_gq),
\end{equation}
where the new functions $\widetilde{Y}_{0C}^{(i)}$ and
$\widetilde{Y}_{CC}^{(i)}$ vanish at $|\omega|=v_gq$. Then the
collision integral (\ref{iq12a}) takes the form
\begin{eqnarray}
\label{iq12d}
&&
\bs{\cal I}'_{12} = - \bs{\cal A}_1 \frac{v_g}{\nu_{01}} \oint q Y_{0A}^{(1)}Y_{00}^{(2)}
+ \bs{\cal A}_2 \frac{v_g}{\nu_{02}} \oint q Y_{00}^{(1)}Y_{0A}^{(2)}
\\
&&
\nonumber\\
&&
\qquad\qquad\qquad
- \bs{\cal C}_1 \frac{v_g}{\nu_{01}} 
\left[\oint q Y_{0C}^{(1)}Y_{00}^{(2)}\theta(|\omega|<v_gq) + 
      \oint q \widetilde{Y}_{0C}^{(1)}Y_{00}^{(2)}\theta(|\omega|>v_gq) \right]
\nonumber\\
&&
\nonumber\\
&&
\qquad\qquad\qquad
+ \bs{\cal C}_2 \frac{v_g}{\nu_{02}} 
\left[\oint q Y_{00}^{(1)}Y_{0C}^{(2)}\theta(|\omega|<v_gq) + 
      \oint q Y_{00}^{(1)}\widetilde{Y}_{0C}^{(2)}\theta(|\omega|>v_gq) \right]
\nonumber\\
&&
\nonumber\\
&&
\qquad\qquad\qquad
- \left(\bs{\cal B}_1 \frac{v_g^2}{\nu_{01}K_1}-\bs{\cal B}_2 \frac{v_g^2}{\nu_{02}K_2}\right) 
\Gamma_0
- \left( \bs{\cal C}_1 \frac{1}{\nu_{01}} - \bs{\cal C}_2 \frac{1}{\nu_{02}}\right)
\Gamma_2, 
\nonumber
\end{eqnarray}
where the last line contains the diverging integrals
\begin{equation}
\label{gamma}
\Gamma_0=\oint q^2Y_{00}^{(1)}Y_{00}^{(2)}, \qquad
\Gamma_2 = 2\oint |\omega|  Y_{00}^{(1)}Y_{00}^{(2)} \theta(|\omega|>v_gq).
\end{equation}

The terms with these diverging rates should be excluded from the
hydrodynamic equations, which reduces the number of independent
macroscopic currents. In order to do so, one has to solve the system
of equations (\ref{eqs12}) for the combinations 
$\bs{\cal B}_1 v_g^2/(\nu_{01}K_1)-\bs{\cal B}_2 v_g^2(\nu_{02}K_2)$ and 
$ \bs{\cal C}_1/\nu_{01} - \bs{\cal C}_2/\nu_{02}$ keeping the
rates $\Gamma_i$ and then take the limit $\Gamma_i\rightarrow\infty$.
This yields the interlayer thermalization conditions
\begin{equation}
\label{trm}
\bs{\cal B}_1 \frac{1}{\nu_{01}K_1}=\bs{\cal B}_2 \frac{1}{\nu_{02}K_2}, \qquad
\bs{\cal C}_1 \frac{1}{\nu_{01}} = \bs{\cal C}_2 \frac{1}{\nu_{02}}.
\end{equation}
At the neutrality point these conditions simplify to 
\begin{equation}
\label{trmdp}
\bs{\cal B}_1(\mu_1=0) = \bs{\cal B}_2(\mu_2=0), \qquad
\bs{\cal C}_1(\mu_1=0) = \bs{\cal C}_2(\mu_2=0).
\end{equation}
Now the number of independent currents and correspondingly the number
of macroscopic equations is reduced from six to four. All terms that
do not contain the diverging rates $\Gamma_i$ can be straightforwardly
simplified using Eqs.~(\ref{trm}). More care is needed when treating
the contributions of the collision integrals (\ref{iq12d}) and
(\ref{ip12a}) where one needs to find the limiting value of the
expressions containing $\Gamma_i$. As a result, we find the thermalized
equations (\ref{eqs12fl}) and (\ref{eqs12dp}). The latter equations
also contain the relaxation rate $\tau_{ss,12}$ is given by
\begin{equation}
\label{tssdp}
\tau^{-1}_{ss,12} = \frac{1}{\nu_0}\oint \left( Y_{00}Y_{CC}-Y_{0C}^2 \right)\theta(|\omega|<v_gq)
+ \frac{1}{\nu_0}\oint 
\left( Y_{00}\widetilde{Y}_{CC}-\widetilde{Y}_{0C}^2 
+4\frac{|\omega|}{v_gq}Y_{00}\widetilde{Y}_{0C}\right)\theta(|\omega|>v_gq),
\end{equation}
appearing from the non-diverging difference between the last two terms in 
Eq.~(\ref{ip12dp}).

\section{Relaxation rates due to electron-phonon interaction}
\label{eph}

\subsection{Electron-phonon collision integral}

Consider the standard form of electron-phonon collision integral. In
graphene it has the following form \cite{dau,hwa,fra,bis,kub,tse,vil,son}:
\begin{subequations}
\label{ieph}
\begin{equation}
{\cal I}_{e-ph} = \sum\limits_1 \left\{f_1 \left[1-f_2\right] {\cal W}_{1\rightarrow 2}
- f_2 \left[1-f_1\right] {\cal W}_{2\rightarrow 1} \right\},
\end{equation}
where
\begin{equation}
\label{wph}
{\cal W}_{1\rightarrow 2} = 2\pi\sum\limits_{\bs{q}} |\lambda_{\bs{v}_1\bs{v}_2}|^2 W_q
\left[\left(1+N_q\right)\delta(\bs{p}_2-\bs{p}_1+\bs{q})\delta(\epsilon_2-\epsilon_1+\omega_q)
+ N_q \delta(\bs{p}_2-\bs{p}_1-\bs{q})\delta(\epsilon_2-\epsilon_1-\omega_q)\right].
\end{equation}
\end{subequations}
Here $N_q$ is the phonon distribution function, $\omega_q$ is the
phonon dispersion, and $W_q$ is the transition matrix element squared. For
acoustic phonons \cite{bis}
\[
W_q = \frac{D^2q^2}{2\rho_m\omega_q},
\]
where $D$ is the screened deformation potential and $\rho_m$ is the
mass density of graphene. At the same time, in graphene inelastic
relaxation may occur through a combined scattering process involving
both a phonon and an impurity \cite{son}. Other possibilities include
two-phonon scattering and phonon-induced intervalley scattering. For
these processes the matrix element is more involved.

We now linearize the collision integral (\ref{ieph}) in the standard
fashion \cite{dau} using Eq.~(\ref{df-ms}) and the similar form of the
non-equilibrium correction to the phonon distribution function
\[
N_q = N_q^{(0)}+\delta N_q, \qquad
\delta N_q = N_q^{(0)} \left(1+N_q^{(0)}\right) \chi = -T \frac{\partial N_q^{(0)}}{\omega_q}\chi.
\]
Consider the first term in Eq.~(\ref{wph}). The same
$\delta$-functions appear also in the second term in Eq.~(\ref{ieph})
describing the reverse process. Combining the two, one finds the
following combination of distribution functions
\[
f_1\left[1-f_2\right]\left(1+N_q\right) - f_2 \left[1-f_1\right] N_q
=
\left[1-f_1\right]\left[1-f_2\right]\left(1+N_q\right)
\left[ \frac{f_1}{1-f_1} - \frac{f_2}{1-f_2} \frac{N_q}{1+N_q}\right].
\]
It is straightforward to check that the expression in square brackets
vanishes in equilibrium. Linearization yields (the non-equilibrium correction
(\ref{habc}) contains the velocity and thus does not contribute to the 
relaxation rates)
\begin{equation}
\label{leph}
f_1\left[1-f_2\right]\left(1+N_q\right) - f_2 \left[1-f_1\right] N_q \approx
f_1^{(0)}\left[1-f_2^{(0)}\right]\left(1+N_q^{(0)}\right)
\left[\delta h_1-\delta h_2 - \chi_q\right].
\end{equation}
The combination of the equilibrium distribution functions in
Eq.~(\ref{leph}) can be further simplified as
\[
f_1^{(0)}\left[1-f_2^{(0)}\right]\left(1+N_q^{(0)}\right)
= - T \frac{\partial N_q^{(0)}}{\omega_q} \left(f_1^{(0)}-f_2^{(0)}\right).
\]
Finally, one may write the linearized electron-phonon collision integral
as a sum of the electron and phonon parts [following Eq.~(\ref{leph})]:
\begin{subequations}
\label{liep}
\begin{equation}
{\cal I}_{e-ph} = {\cal I}_e + {\cal I}_{ph},
\end{equation}
where the electronic part is given by
\begin{eqnarray}
&&
{\cal I}_e = \frac{\pi}{4} \sum_{\bs{q}} \frac{W_q}{\sinh^2(\omega_q/2T)} 
\sum\limits_1 |\lambda_{\bs{v}_1\bs{v}_2}|^2
\left[
\delta(\bs{p}_2-\bs{p}_1+\bs{q})\delta(\epsilon_2-\epsilon_1+\omega_q)
- \delta(\bs{p}_2-\bs{p}_1-\bs{q})\delta(\epsilon_2-\epsilon_1-\omega_q)
\right]
\nonumber
\\
&&
\nonumber\\
&&
\qquad\qquad\qquad\qquad\qquad
\qquad\qquad
\times
\left(\tanh\frac{\epsilon_2-\mu}{2T}-\tanh\frac{\epsilon_1-\mu}{2T}\right)
\left[\delta h_1-\delta h_2\right].
\end{eqnarray}
\end{subequations}
In this paper we consider the phonon system to be at equilibrium and
therefore neglect the phonon part of the collision integral. This
means that all back-action effects, such as phonon drag, are
neglected. For some physical processes, most notably, thermoelectric
effects, such processes might be important. Then one has to consider
the phonon kinetic equation on equal footing with Eq.~(\ref{beq0}).

\subsection{Energy relaxation rates}

The relaxation rates are obtained by integrating the collision
integral (\ref{liep}). The ``energy'' continuity equation is obtained
by multiplying the kinetic equation by $\epsilon$ and integrating over
all states. The corresponding integrated collision integral has the
form
\begin{eqnarray}
&&
\sum_{2}\epsilon_2{\cal I}_e = \frac{\pi}{4} \sum_{\bs{q}} \frac{W_q}{\sinh^2(\omega_q/2T)} 
\sum\limits_{1,2} \epsilon_2 |\lambda_{\bs{v}_1\bs{v}_2}|^2
\left[
\delta(\bs{p}_2-\bs{p}_1+\bs{q})\delta(\epsilon_2-\epsilon_1+\omega_q)
- \delta(\bs{p}_2-\bs{p}_1-\bs{q})\delta(\epsilon_2-\epsilon_1-\omega_q)
\right]
\nonumber
\\
&&
\nonumber\\
&&
\qquad\qquad\qquad\qquad\qquad
\qquad\qquad
\times
\left(\tanh\frac{\epsilon_2-\mu}{2T}-\tanh\frac{\epsilon_1-\mu}{2T}\right)
\left[\delta h_1-\delta h_2\right].
\end{eqnarray}
The difference between the non-equilibrium distribution functions reads
\[
\delta h_1-\delta h_2 = \frac{1}{e\nu_0T}
\left[\mathfrak{b} \frac{\epsilon_1-\epsilon_2}{K} + \mathfrak{c}
\left[{\rm sign}(\epsilon_1)-{\rm sign}(\epsilon_2)\right] \right].
\]
Consequently, we can define two relaxation rates
\begin{equation}
\label{er}
\sum_{2}\epsilon_2{\cal I}_e = - \frac{\mathfrak{b}}{\tau_{Eb}} + \frac{\mathfrak{c}}{\tau_{Ec}}.
\end{equation}
Specifically at the neutrality point we can use Eqs.~(\ref{saim0}) and
express the integrated collision integral in terms of the energy and
imbalance densities
\begin{equation}
\label{er0}
\sum_{2}\epsilon_2{\cal I}_e = \frac{e\delta u}{K\Delta(0)} \left[\frac{1}{\tau_{Eb}}
+ \frac{\pi^2}{12\ln^22}\frac{1}{\tau_{Ec}}\right]
-
\frac{e\delta\rho}{\Delta(0)} \left[ \frac{{\cal N}_2(0)}{\tau_{Ec}} + 
\frac{\pi^2}{12\ln^22}\frac{1}{\tau_{Eb}}\right].
\end{equation}

\subsection{Imbalance relaxation rates}

Similarly, we find the imbalance relaxation rates. The corresponding integrated 
collision integral has the form
\begin{eqnarray}
&&
\sum_{2}{\rm sign}(\epsilon_2){\cal I}_e = \frac{\pi}{4} \sum_{\bs{q}} 
\frac{W_q}{\sinh^2(\omega_q/2T)} 
\sum\limits_{1,2} {\rm sign}(\epsilon_2) |\lambda_{\bs{v}_1\bs{v}_2}|^2
\left(\tanh\frac{\epsilon_2-\mu}{2T}-\tanh\frac{\epsilon_1-\mu}{2T}\right)
\\
&&
\nonumber\\
&&
\qquad\qquad\qquad\qquad\qquad
\times
\left[
\delta(\bs{p}_2-\bs{p}_1+\bs{q})\delta(\epsilon_2-\epsilon_1+\omega_q)
- \delta(\bs{p}_2-\bs{p}_1-\bs{q})\delta(\epsilon_2-\epsilon_1-\omega_q)
\right]
\left[\delta h_1-\delta h_2\right].
\nonumber
\end{eqnarray}
Clearly, only inter-band scattering processes contribute to this
collision integral (unlike the case of the energy relaxation, where
both inter- and intra-band processes have to be taken into account).

For general doping we define the following relaxation rates
\begin{equation}
\label{ir}
\sum_{2}{\rm sign}(\epsilon_2){\cal I}_e 
= \frac{\mathfrak{b}}{\tau_{Ib}} - \frac{\mathfrak{c}}{\tau_{Ic}},
\end{equation}
where
\begin{equation}
\tau_{Ib} = \tau_{Ec}.
\end{equation}
At the neutrality point this yields
\begin{equation}
\label{ir0}
\sum_{2}{\rm sign}(\epsilon_2){\cal I}_e = 
-\frac{e\delta u}{K\Delta(0)} \left[\frac{1}{\tau_{Ib}}
+ \frac{\pi^2}{12\ln^22}\frac{1}{\tau_{Ic}}\right]
+
\frac{e\delta\rho}{\Delta(0)} \left[ \frac{{\cal N}_2(0)}{\tau_{Ic}} + 
\frac{\pi^2}{12\ln^22}\frac{1}{\tau_{Ib}}\right].
\end{equation}

Combining the above electron-phonon collision integrals into the two
continuity equations for the energy and imbalance densities, we find
Eqs.~(\ref{ce0}), where the matrix matrix elements of ${\cal T}_{ph}$
combine the above relaxation rates. The rates $\tau_{Ec}^{-1}$ and
$\tau_{Ic}^{-1}$ are determined by the interband scattering processes
in contrast to the rate $\tau_{Eb}^{-1}$ which contains contribution of 
the intraband processes as well. Therefore,  
\begin{equation}
\tau_{Eb} \ll \tau_{Ec} \leqslant \tau_{Ic},
\end{equation}
such that the matrix ${\cal T}_{ph}$ has two positive eigenvalues as
it should.

\section{Continuity equations in double-layer systems}
\label{ad}

Electron-electron interaction does not contribute to continuity
equations in monolayer graphene (\ref{ceqsdp}) due to the conservation
laws. In double-layer systems, only the electric charge is conserved
leaving the corresponding continuity equation trivial
[cf. Eq.~(\ref{jceqdp})], while the quasiparticle energy and imbalance
are affected by interlayer scattering.

\subsection{Energy relaxation due to electron-electron interaction}

The continuity equation for energy is obtained by multiplying the
kinetic equation by $\epsilon$ and integrating over all states.
Integrating the collision integral that describes interlayer
electron-electron interaction we find [cf. Eq.~(\ref{iq12})]
\begin{eqnarray}
\label{ceq12}
&&
{\cal I}'_{12} = - \frac{1}{32} \int \frac{d^2q d\omega}{(2\pi)^3}
\frac{|U_{12}(\bs{q},\omega)|^2}{\sinh^2(\omega/2T)}
\\
&&
\nonumber\\
&&
\qquad\qquad\qquad
\times
\sum_{1, 1'} 
(2\pi)^3 |\lambda_{\bs{v}\bs{v}'}|^2
\delta(\epsilon_1 - \epsilon_1' + \omega)
\delta(\bs{p}_1 - \bs{p}_1' + \bs{q})
\left[\tanh\frac{\epsilon_1-\mu_2}{2T} - \tanh\frac{\epsilon_1+\omega-\mu_2}{2T}\right]
\nonumber\\
&&
\nonumber\\
&&
\qquad\qquad\qquad
\times
\sum_{2,2'} (\epsilon_2-\epsilon_2')
(2\pi)^3 |\lambda_{\bs{v}\bs{v}'}|^2
\delta(\epsilon_2 - \epsilon_2' - \omega)
\delta(\bs{p}_2 - \bs{p}_2' - \bs{q})
\left[\tanh\frac{\epsilon_2-\mu_1}{2T} - \tanh\frac{\epsilon_2-\omega-\mu_1}{2T}\right]
\nonumber\\
&&
\nonumber\\
&&
\qquad\qquad\qquad
\times
\left[ \delta h_{2,1'} + \delta h_{1,2'} - \delta h_{2,1} - \delta h_{1,2} \right].
\nonumber
\end{eqnarray}
Using the explicit form of the distribution function (\ref{dhabc}) and
energy conservation we find
\[
\delta h_{2,1'} - \delta h_{2,1} = \frac{1}{e\nu_{02}T}\left[
\mathfrak{b}_2\frac{\epsilon_1'-\epsilon_1}{K_2} +
\mathfrak{c}_2\left({\rm sign}(\epsilon_1')-{\rm sign})\epsilon_1)\right)
\right]
=
\frac{1}{e\nu_{02}T}\left[
\mathfrak{b}_2\frac{\omega}{K_2} +
\mathfrak{c}_2{\rm sign}(\omega)
\right],
\]
and similarly for the first layer. As a result
\begin{equation}
\label{ceq12-2}
e{\cal I}'_{12} = - \left[\frac{\mathfrak{b}_1}{\nu_{01}K_1} - 
\frac{\mathfrak{b}_2}{\nu_{02}K_2}\right]
\oint \omega^2 Y_{00}^{(1)} Y_{00}^{(2)}
-
\left[\frac{\mathfrak{c}_1}{\nu_{01}} - 
\frac{\mathfrak{c}_2}{\nu_{02}}\right]
\oint |\omega| Y_{00}^{(1)} Y_{00}^{(2)}.
\end{equation}

\subsection{Energy relaxation due to electron-electron interaction}

Similarly to the previous Section, we find the contribution 
of electron-electron interaction to the continuity equation
for quasiparticle imbalance
[cf. Eq.~(\ref{ip12})]
\begin{eqnarray}
\label{cep12}
&&
{\cal I}''_{12} = - \frac{1}{32} \int \frac{d^2q d\omega}{(2\pi)^3}
\frac{|U_{12}(\bs{q},\omega)|^2}{\sinh^2(\omega/2T)}
\\
&&
\nonumber\\
&&
\qquad\qquad
\times
\sum_{1, 1'} 
(2\pi)^3 |\lambda_{\bs{v}\bs{v}'}|^2
\delta(\epsilon_1 - \epsilon_1' + \omega)
\delta(\bs{p}_1 - \bs{p}_1' + \bs{q})
\left[\tanh\frac{\epsilon_1-\mu_2}{2T} - \tanh\frac{\epsilon_1+\omega-\mu_2}{2T}\right]
\nonumber\\
&&
\nonumber\\
&&
\qquad\qquad
\times
\sum_{2,2'} \left[{\rm sign}(\epsilon_2)-{\rm sign}(\epsilon_2')\right]
(2\pi)^3 |\lambda_{\bs{v}\bs{v}'}|^2
\delta(\epsilon_2 - \epsilon_2' - \omega)
\delta(\bs{p}_2 - \bs{p}_2' - \bs{q})
\left[\tanh\frac{\epsilon_2-\mu_1}{2T} - \tanh\frac{\epsilon_2-\omega-\mu_1}{2T}\right]
\nonumber\\
&&
\nonumber\\
&&
\qquad\qquad
\times
\left[ \delta h_{2,1'} + \delta h_{1,2'} - \delta h_{2,1} - \delta h_{1,2} \right].
\nonumber
\end{eqnarray}
\end{widetext}
Using the explicit form of the distribution function (\ref{dhabc}) and
energy conservation we find
\begin{eqnarray}
\label{cep12-2}
&&
e{\cal I}''_{12} = - \left[\frac{\mathfrak{b}_1}{\nu_{01}K_1} - 
\frac{\mathfrak{b}_2}{\nu_{02}K_2}\right]
\oint |\omega| Y_{00}^{(1)} Y_{00}^{(2)}
\\
&&
\qquad\qquad\qquad
-\left[\frac{\mathfrak{c}_1}{\nu_{01}} - 
\frac{\mathfrak{c}_2}{\nu_{02}}\right]
\oint Y_{00}^{(1)} Y_{00}^{(2)}.
\nonumber
\end{eqnarray}

\subsection{Thermalization in finite-size samples}

The collision integrals (\ref{ceq12-2}) and (\ref{cep12-2}) contain
formally diverging expressions [similar to Eqs.~(\ref{gamma})]:
\begin{equation}
\label{gn}
\oint \omega^2 Y_{00}^{(1)} Y_{00}^{(2)}, \quad
\oint |\omega| Y_{00}^{(1)} Y_{00}^{(2)}, \quad
\oint Y_{00}^{(1)} Y_{00}^{(2)}.
\end{equation}

If one assumes equal strength of intra- and interlayer Coulomb
interaction, then one needs to perform the interlayer thermalization
procedure, described in Appendix~\ref{it}. In finite-size systems,
this procedure has to include the continuity equations containing the
formally diverging terms (\ref{gn}). Since the macroscopic
equations contain gradient terms, the resulting hydrodynamic equations
will now contain gradients of the driving current $\bs{j_1}(y)$.

On the other hand, at the phenomenological level one may assume the
interlayer interaction to be weaker than the intralayer
interaction. In that case, the latter is responsible for forming the
hydrodynamic modes, while the former [where the terms (\ref{gn}) are
  treated as finite] play the role of additional relaxation rates.
This way one obtains the phenomenological model of
Ref.~\onlinecite{meg}, which qualitatively captures the essential
physics of the system.

\end{document}